\documentclass[useAMS,usenatbib]{mnras}
\usepackage{graphicx,amssymb,url,enumerate}
\usepackage{amsmath}
\usepackage{gensymb}

\def\kms{km~s$^{-1}$}
\def\kmsp{km~s$^{-1}$~kpc}
\def\dir{./plots/} 

\title[Phase Wrapping in the Wobbly Galaxy]{Phase Wrapping of Epicyclic Perturbations in the Wobbly Galaxy}
\author[de la Vega et al.]{
Alexander de la Vega$^{1,2}$, 
Alice C. Quillen$^2$,
Jeffrey L. Carlin$^{3,4}$,
\newauthor
Sukanya Chakrabarti$^5$
and Elena D'Onghia$^{6,7}$
\\
$^1$ Department of Physics and Astronomy, Johns Hopkins University, Baltimore, MD 21218, USA\\
$^2$ Department of Physics and Astronomy, University of Rochester, Rochester, NY 14627, USA\\
$^3$ Department of Physics, Applied Physics and Astronomy, Rensselaer Polytechnic Institute, Troy, NY 12180, USA \\
$^4$ Department of Physics and Astronomy, Earlham College, Richmond, IN 47374, USA \\
$^5$ School of Physics and Astronomy, Rochester Institute of Technology, 84 Lomb Memorial Drive,
Rochester, NY 14623 \\
$^6$ Department of Astronomy,    University    of    Wisconsin-Madison, Madison, WI 53706, USA \\
$^7$ Alfred P. Sloan Fellow \\
}

\begin{document}
\maketitle

\begin{abstract}
We use test-particle integrations to show that epicyclic motions excited by a pericentre passage of a dwarf
galaxy  could account for bulk vertical velocity streaming motions
recently observed in the Galactic stellar disc near the Sun. 
We use fixed potential test-particle integrations to isolate the role of phase wrapping of epicyclic perturbations
from bending and breathing waves or modes, which require self-gravity to oscillate. 
Perturbations from  a fairly massive Sagittarius dwarf galaxy, $M_d \sim 2.5 \times 10^{10}M_\odot$, 
are required to account for the size  of the observed streaming motions from its orbital pericentre
approximately a Gyr ago. A previous passage of the dwarf through the Galactic disc approximately 2.2 Gyr ago
(with a then more massive dwarf galaxy) is less effective. 
If phase wrapping of epicyclic perturbations is responsible for stellar streaming motions in the Galactic disc, 
then there should be variations in velocity gradients on scales of a few kpc in
the vicinity of the Sun.  

 
\end{abstract}


\section{Introduction}

Significant bulk motions of stars, or streaming velocities, have recently been detected in large scale stellar surveys 
such as the RAdial Velocity Experiment (RAVE, \citealt{rave}) and 
Large Area Multi-Object Spectroscopic Telescope (LAMOST, \citealt{lamost}) 
radial velocity surveys \citep{siebert11,williams13,carlin13,xu15,sun15}.
Vertical wavelike structures in the stellar density distribution are also seen in the Sloan Digital Sky Survey (SDSS) data \citep{widrow12,xu15}.
Curiously, the bulk motions of stars vary as a function of height above
the plane, and the  vertical bulk motions exhibit patterns of compression and rarefaction.
The observed vertical velocity gradients could be due to 
 heating from internal perturbations, such as spiral arms or bars \citep{faure14, monari15}, 
 or bending and breathing waves or modes \citep{widrow12,gomez13} that
 could have been excited by a dwarf galaxy \citep{widrow14}.
 Vertical density and velocity structures  arise in simulated Milky Way discs  perturbed  
 by a Sagittarius sized dwarf galaxy \citep{gomez13} that could also have induced
 warped and ringed structures (such as the Monocerous Ring; \citealt{newberg02}) 
 in the outer Galaxy (e.g., \citealt{kazant08,younger08,quillen09,purcell11}).  
 
Pericentre approaches and passages through the Galactic disc by  
the Sagittarius dwarf galaxy would have excited both vertical and radial epicyclic motions in stars. 
A perturbation in the disc could affect a localised region where stars are pushed to the same phase in their epicycles.
Over time, these perturbed stars would see a large spread in phases, called phase wrapping, 
due to the dependence of vertical and radial (epicyclic) oscillation frequencies on orbital angular momentum, eccentricity
and inclination 
(e.g., \citealt{minchev09}), giving spiral and warped structures in the disc \citep{quillen09}. 
 However, the perturbation of a dwarf galaxy could also excite
vertical bending and breathing waves or modes in the disc \citep{weinberg91,gomez13,widrow14, widrow15}.
N-body simulations would be expected to give a more realistic simulation of perturbations to the Galactic disc
caused by the Sagittarius dwarf galaxy. However, with N-body simulations it is difficult to
differentiate between excitation of bending or breathing waves or modes from phase wrapping of epicyclic motions.
Waves and vibration modes are a property of a self-gravitating disc \citep{widrow12}, and so would not be present
in a non-interacting test-particle simulation where particle motions are integrated in a 
fixed Galactic potential. Here we use  
test particle integrations to isolate and study the role of epicyclic phase wrapping and explore which
structures in the local velocity field might be a result of vertical and radial epicyclic motions excited in the disc
by a previous pericentre or a previous passage through the Galactic disc of the Sagittarius dwarf galaxy.

\section{Test-Particle Integrations}

We integrate particle orbits for two different sets of initial conditions using a fixed gravitational potential for the Milky Way.
Particle orbits are integrated using the python library \texttt{galpy} \citep{bovy15}
using the \texttt{MWPotential2014} model gravitational potential.
This potential is designed to give  a realistic model for the Milky Way and is described in detail
by \citet{bovy15} in his section 3.5.
Our procedure is as follows:  We generate a thin disc of $10^7$ test particles.  
We perturb the velocities of each particle, instantaneously approximating
perturbations caused by an encounter from a dwarf galaxy.   
We then integrate the particles to the present day in the static Galaxy potential alone.
Each set of integrations required 3--5 days of computations on a 
Mac-Mini computer with a 2.4 GHz IntelCore 2 Duo processor.
The perturbations are applied instantaneously,
rather than directly integrated as a function of time,
so as to separate between the excitations caused by the perturbations and subsequent phase wrapping in the fixed background
Galactic potential.

We consider the role of two perturbations from the dwarf galaxy, separately,
so that we can contrast and compare their different roles in exciting epicyclic motions in the stars. 
We run two separate integrations, one beginning with perturbations caused by the passage of the dwarf galaxy through
the Galactic plane approximately 2.2 Gyr ago, and the second integration, beginning with perturbations 
caused by the dwarf galaxy's pericentre approximately 1.1 Gyr ago.

\begin{figure}
\includegraphics[width=3.2in]{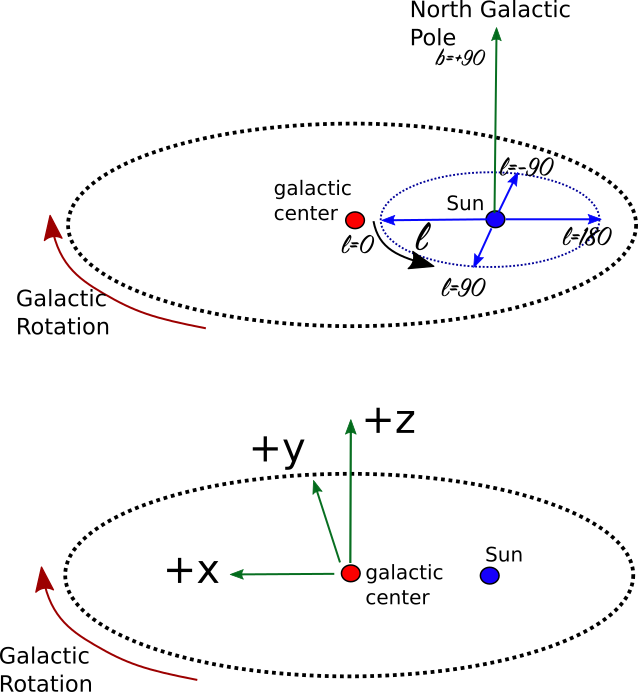} 
\caption{The top panel shows Galactic latitude and longitude (on the sky). The bottom panel shows the Galactocentric 
cartesian coordinate system. 
\label{fig:coord}
}
\end{figure}

We work in the right-handed Galactocentric coordinate  system used by \citet{law10}. 
The relation between Galactic latitude and longitude (on the sky 
and with origin at the Sun)
and the Cartesian Galactocentric coordinate system with origin at the Galactic Center is illustrated in Figure \ref{fig:coord}.
This coordinate system gives a current
location of the Sagittarius dwarf at $X,Y,Z= (19.0, 2.7,-6.9)$ in kpc with a distance from the Sun of $\sim 28$ kpc and 
Galactocentric solar radius $R_\odot=8.0$ kpc. 
In this coordinate system the current location of the Sun is $(-R_\odot,0,0)$ and this differs
from that used by \citet{carlin13} and \citet{sun15} who place the Sun at $(R_\odot, 0, 0)$.
We also use Galactocentric cylindrical coordinates $(R, \Theta, Z)$, with Galactic rotation $\dot \Theta >0$, 
and in these the location
of the Sun is $(R_\odot, \Theta_\odot ,0)$ with $\Theta_\odot = \pi$, differing from  
\citet{carlin13} and \citet{sun15} who adopt $\Theta_\odot = 0$.
We work in units of kpc, $M_\odot$ and velocities are given in \kms.
The adopted circular velocity at $R_\odot$ is $V_\odot = 220$~ \kms
and  the period of Galactic rotation at $R_\odot$ is $P=2\pi R_\odot/V_\odot =$ 0.23 Gyr.


\subsection{Thin disc prior to perturbation by the dwarf galaxy}

Prior to perturbation by a dwarf galaxy encounter, 
particles are initially evenly distributed in a thin disc.   
Working in Galactocentric cylindrical coordinates 
 we start by choosing azimuthal angle, $\Theta$, from a uniform
distribution $0 \leq \Theta \leq 2 \pi$. Guiding radii are chosen to be uniformly distributed
 $5 \leq R_g \leq 30$ kpc, giving a surface density proportional to $1/R$.  This has a shallower radial gradient
 than an exponential distribution
 but allows us to more efficiently study the structure of the outer Galaxy by increasing the number of test
 particles in that region.
 
After the guiding radius $R_g$ is chosen, the initial radius, radial and tangential velocity components are chosen using a first order 
(in radial action variable) epicyclic approximation.
The epicyclic amplitudes $a_r$ are chosen to be uniformly distributed between zero and $a_{rmax}$.  Epicyclic angles, $\phi_r$, are chosen from 
a uniform distribution with $0 \leq \phi_r \leq 2\pi$.  Initial radii and radial velocities are set to 
\begin{eqnarray}
 R &=& a_r \sin( \phi_r) + R_g \nonumber \\
V_R &=& a_r \kappa(R_g) \cos(\phi_r). \label{eqn:epi}
\end{eqnarray} 
The vertical component of the orbit's angular momentum  is $L_Z=R_g V_c(R_g)$ 
and the initial tangential velocity component computed as
$V_\theta =  L_Z/R$.
Here $V_c(R_g)$ is the rotation curve and $\kappa(R_g)$ is the radial epicyclic frequency, and these are computed 
using the galactic potential \texttt{MWPotential2014} using the \texttt{vcirc} and \texttt{epifreq} routines from \texttt{galpy}.
 The initial vertical positions and velocities are similarly chosen by choosing an epicyclic amplitude, $a_z$, uniformly distributed
 between zero and $a_{zmax}$, and   vertical epicyclic angle $\phi_z$, uniformly distributed between zero and $2 \pi$, 
 then setting initial $Z$ and vertical velocity component $V_Z$ to be 
\begin{eqnarray}
 Z &=& a_z \sin( \phi_z)  \nonumber \\
V_Z &=& a_z \nu(R_g) \cos(\phi_z). \label{eqn:vert_epi}
\end{eqnarray}   
Here $\nu(R_g)$ is the vertical oscillation frequency, which is computed in the \texttt{MWPotential2014} galactic
potential using the \texttt{verticalfreq} routine from \texttt{galpy}. 
Setting $a_{rmax} = 1.6$ kpc and $a_{zmax} = 0.4$ kpc, the resulting disc has velocity dispersions
$\sigma_R \approx 20$ \kms, $\sigma_Z \approx 10$ \kms, and $\sigma_\phi \approx 15$ \kms \ at $R=R_\odot$
that are consistent with thin disc values computed in the Solar neighbourhood (comparing to Figure 31 by \citealt{nordstrom04}). 
The parameters $a_{rmax}$ and $a_{zmax}$ are independent of galactocentric guiding radius, ($R_g$), corresponding
to a thin disc with scale height independent of radius.  As $\nu/\kappa$ varies from 2.0 at $R_\odot$ to 1.56 at 25 kpc
the ratio $\sigma_Z/\sigma_R$ is 0.5 at $R_\odot$ and drops to 0.4 at 25 kpc.
The expected vertical density profile $\rho(z)$ is logarithmically sensitive to $z/a_{zmax}$ with a cusp at the midplane.
From the initial stellar distribution, we numerically measured the vertical dispersion of the initial density profile at $R_\odot$ finding 
a width $\sqrt{\langle z^2 \rangle} = 163$ pc.
We also numerically measured the asymmetric drift $v_a = \langle V_c - V_\theta \rangle  \sim 3 $ \kms at $R_\odot$.

\subsection{Perturbations from a Dwarf Galaxy}

\begin{figure}
\centering\includegraphics[width=3.3in]{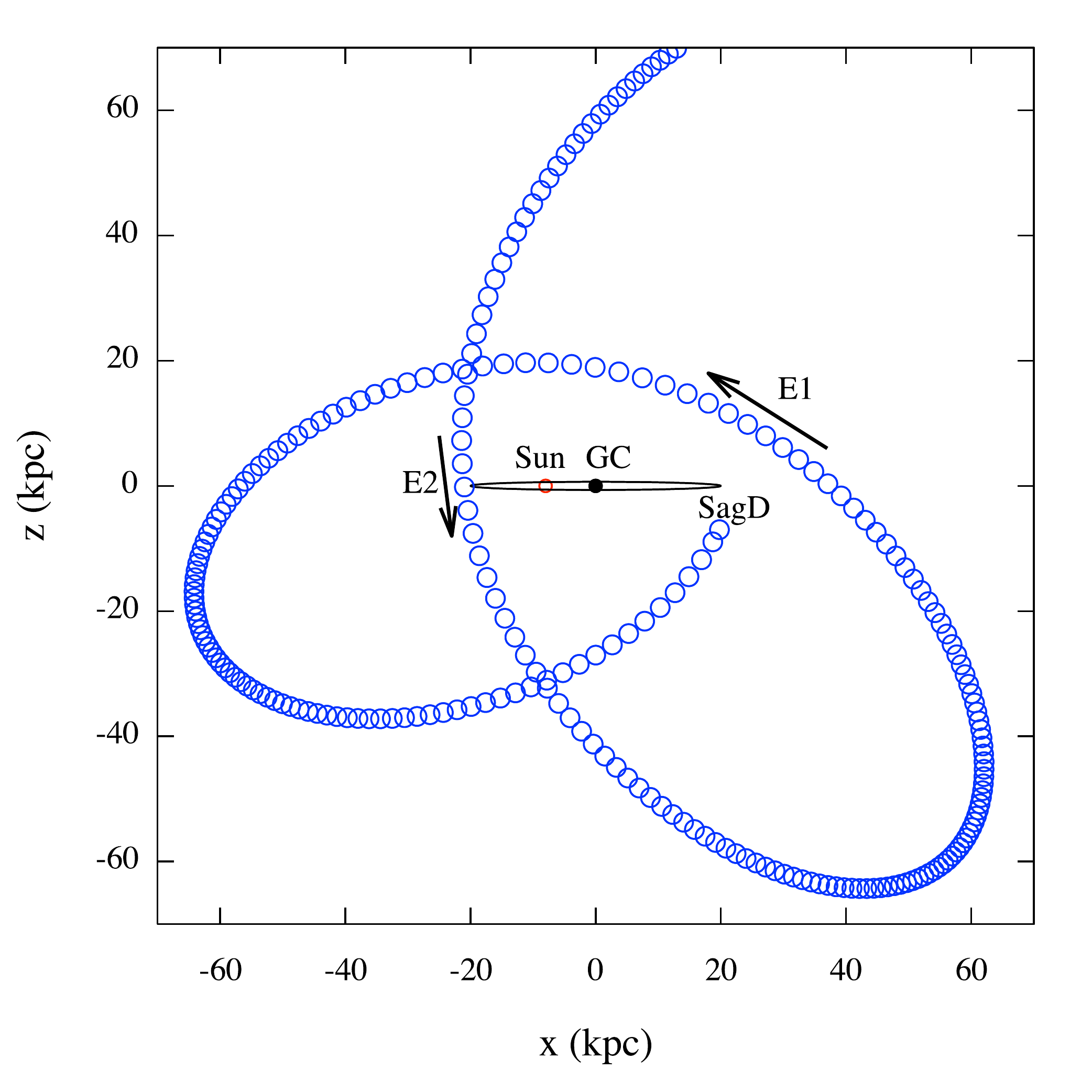}
\caption{Illustrating the Sagittarius dwarf orbit from the E model by \citet{chakra14}. The current position of the dwarf in its orbit is 
labelled "SagD" and the passage through the galactic plane 2.2 Gyr ago labelled with E2 and a black arrow. The previous pericentre, 
1.1 Gyr ago, is labelled E1 and shown with another black arrow. Axes are in Galactocentric coordinates and in units of kpc.
\label{fig:orbit}
}
\end{figure}

The orbit of the Sagittarius dwarf galaxy has been inferred from modeling the continuous stream of debris 
in the Sagittarius dwarf galaxy's tidal tails (e.g., \citealt{johnston05,law10,purcell11}) and resembles a trefoil knot 
(see Figure 1 by \citealt{johnston05} and our illustration in Figure \ref{fig:orbit}).
The Sagittarius dwarf galaxy nucleus is currently near pericentre in its orbit.    
The Sagittarius dwarf galaxy nucleus previously passed through the Galactic plane
about 0.4 Gyr ago, but at large Galactocentric radius $R \gtrsim 50$ kpc.   The previous pericentre occurred
about 1 Gyr ago and at positive $Z$ in our Galactocentric coordinate system (see lower left panel
of Figure 8 by \citealt{law10}).  Our  illustration in Figure \ref{fig:orbit} labels the last pericentre as E1.
About 2 Gyr ago, the dwarf galaxy's nucleus passed through the Galactic plane when it was nearly at pericentre.
This event is labelled as E2 in our illustration.
In Figure \ref{fig:orbit} we plot a model orbit by \citet{chakra14} (that labelled E in their Table 1), that matches the observed proper
motions for stars in the Sagittarius dwarf galaxy nucleus.  Using this orbit,
we list in Table \ref{tab:dwarf} the positions and velocities of the dwarf galaxy nucleus at the E1 pericentre and the E2 passage
through the Galactic plane.  These two events could have caused large perturbations to the disc stars
and we use them to generate velocity perturbations for our test-particle integrations.

\begin{table*}
    \centering
    \caption{Parameters for the two dwarf galaxy encounters labelled in Figure \ref{fig:orbit}}
    \begin{tabular}{ccccc}
    \hline
  Encounter & $t \ \rm{(Gyr  \ ago)}$ & Mass $ \left( \times 10^{10} \ M_{\odot} \right)$ & \textbf{v} $\left( \rm{km \ s} ^{-1} \right)$ & \textbf{r} (kpc)  \\
    \hline
  E1 & 1.1 & 2.5 & $\left( -339, -44, 76 \right)$ & $\left( 4, 8, 18 \right)$ \\
  E2 & 2.2 & 5.0 & $\left( 4, -111, -337 \right)$ & $\left( -21, -4, 0 \right)$ \\
      \hline
    \end{tabular}
    \label{tab:dwarf}
\end{table*}

After generating a thin disc we use an instantaneous hyperbolic orbit approximation (often used
to derive dynamical friction and gravitational heating rates, see \citealt{B+T} section 7.1)
 to perturb the velocities of each particle.
Two sets of integrations are done, each using a separate encounter listed in Table \ref{tab:dwarf}.   
We use a hyperbolic orbit to compute the velocity perturbations on each particle (applied instantaneously at the beginning of the orbital integration).  
Our procedure for computing the velocity
perturbation from the dwarf encounter positions and velocities in Table \ref{tab:dwarf} is described in more detail in our appendix \ref{ap:hyper}.
Using this approximation the velocity perturbation $\Delta {\bf v}$ is proportional to the mass of the dwarf galaxy, $M_d$.
Hence, perturbations from larger or smaller dwarf galaxy masses can be estimated by scaling the
resulting velocity perturbations.

The progenitor dwarf galaxy mass for the E model by \citet{chakra14} is $M_d = 10^{10} M_\odot$ and lighter than both the `light' 
($3 \times 10^{10}M_\odot$) and `heavy' ($10^{11} M_\odot$)
Sagittarius dwarf galaxy masses explored by \citet{purcell11}. 
\citet{purcell11,chakra14} take into account dynamical friction and
tidal stripping of the dwarf galaxy. The dwarf galaxy mass is similar to its initial mass until pericentre near E2, approximately
2 Gyr ago, at which time the dwarf loses about half of its mass (see Figure S2 by \citealt{purcell11}). 
The dwarf galaxy mass then remains constant until the E1 pericentre.
Consequently, we take the mass of the dwarf to be equal to its progenitor mass for the E2 encounter and half that at the E1 encounter.
For the E2 encounter we use a dwarf galaxy mass of $M_{d_{E2}} = 5 \times 10^{10} M_\odot$ (which implies $M_{d_{E1}} = 2.5 \times 10^{10} M_\odot$), in between the light and
heavy models by \citet{purcell11}, but heavier than the E model by \citet{chakra14}.

\subsection{Initial mean velocities}
\label{sec:init}

\begin{figure*}
\includegraphics[ width=5.5in]{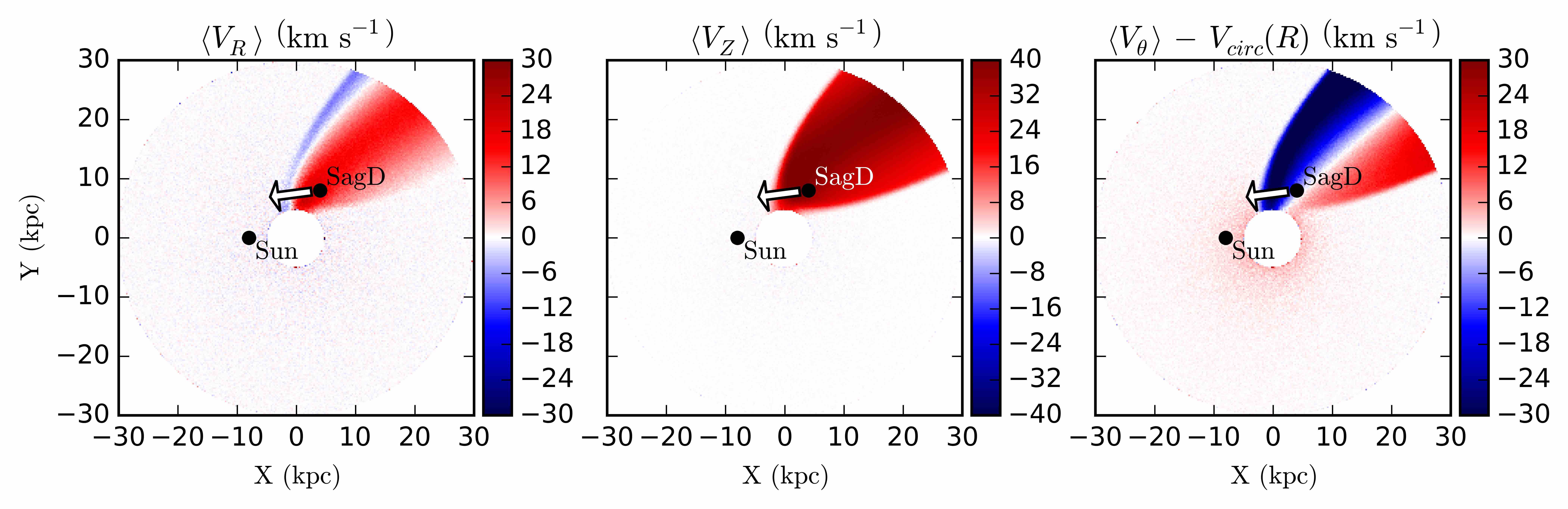}
\includegraphics[ width=5.5in]{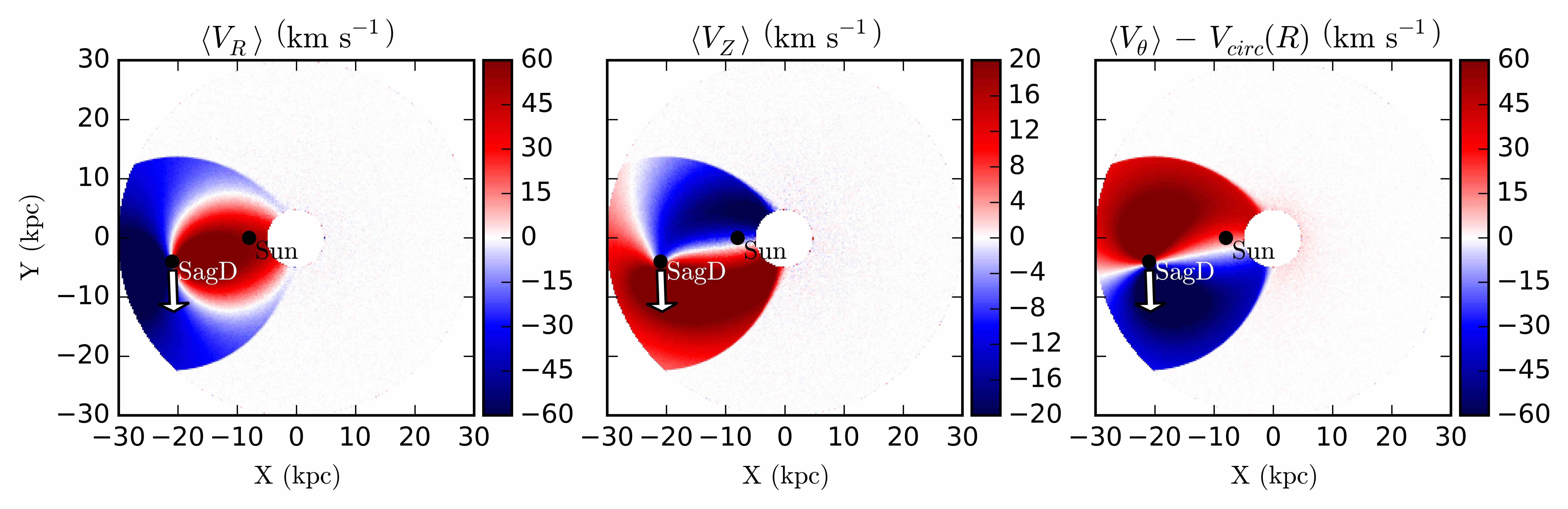}
\caption{Mean particle velocity distributions as a function of $X,Y$  in the Galaxy prior to orbital integration.
The initial thin discs have been instantaneously  perturbed using an instantaneous hyperbolic orbit approximation
(described in appendix \ref{ap:hyper}).  In the top panels, we show the mean velocities induced
by the pericentre encounter 1 Gyr ago (E1 encounter).
The top left panel shows
the mean radial velocity component, the top middle panel, the mean vertical velocity component
and the top right panel the mean tangential velocity component subtracted by $V_{circ}(R)$. Each panel displays  the position and direction of the Sagittarius Dwarf in $X, Y$ with a black dot and an arrow.  
For scale the current position of the Sun is also shown, however the Sun would have been at a different azimuthal angle
during the encounter.  Galactic rotation is counter-clockwise.
On the bottom a similar set of panels show mean velocities but for the passage through the Galactic plane 2 Gyr ago (E2 encounter).
Parameters for the two encounters are listed in Table \ref{tab:dwarf}. 
Nearby particles differ in vertical epicyclic angles by $\pi$ for the E2 encounter, whereas they all
have the same phase in the E1 encounter. Even though the dwarf mass is lower for the E1 encounter, it is more effective
at exciting vertical motions.
}
\label{fig:subpini}
\end{figure*}

We generate 10 million particles for the E1 and the E2 encounters, separately. Following generation of the thin disc and the
instantaneous velocity perturbation due to the E1 or E2 encounters, (and prior to orbit integration) 
we compute mean velocities in 0.25 kpc square bins
in $X$ and $Y$, summing together the velocity components for all particles within each bin.   
Figure \ref{fig:subpini} shows the initial mean radial velocity component, $\langle V_R \rangle$,
the mean vertical velocity component  $\langle V_Z \rangle$, and the mean tangential velocity component subtracted
by the circular velocity as a function of radius, $\langle V_\theta \rangle - V_{circ}(R)$, for the E1 and E2 encounters.
The E2 encounter has velocity vector nearly but not exactly perpendicular to the Galaxy midplane.
The velocity vector is tilted slightly so that the orbit is at positive $Z$ above the midplane at negative $Y$
 and vice-versa at positive $Y$. The tilt of the dwarf's orbit
gives a negative $V_Z$ perturbation to the particles in the disc at positive $Y$ 
and vice-versa at negative $Y$. We see this as nearby red and blue regions in the $\langle V_Z\rangle$ panel
in Figure \ref{fig:subpini}. These nearby particles differ in vertical epicycle angle, $\phi_z$,  by $\pi$ 
(see equations \ref{eqn:vert_epi}). 
The situation is different for the E1 encounter. Because the dwarf galaxy passes above the Galactic plane in this encounter,
both $\langle V_R \rangle$ and $\langle V_Z \rangle$ are in the same direction, though $\left<V_\theta \right>$ has both positive and negative
perturbations (see top panels in Figure \ref{fig:subpini}).
Because $\left<V_Z\right>$ is positive the perturbed particles have the same phase $\phi_z \sim 0$
and the size of $\langle V_Z \rangle$ gives vertical epicyclic amplitude $a_z \sim 0.05$ in units of radius (using a typical $V_Z$ and equations \ref{eqn:vert_epi}).
The sign change in the $V_\theta$ perturbation gives epicyclic angle $\phi_r \sim \pi/2$ where $\left<V_\theta \right>$ is lower than the circular velocity
and $\phi_r \sim -\pi/2$ where $\left< V_\theta \right>$ exceeds the circular velocity (using equation \ref{eqn:epi}) and epicyclic amplitude 
$a_r \sim 0.15$ in units of radius.
Even though the dwarf mass used for the E1 encounter is half of that in the E2 encounter, the
$V_Z$ perturbation is larger. This is because the E1 encounter has its closest approach above 
rather than in the Galactic plane. Even though the dwarf mass is lower for the E1 encounter, it is more
effective at exciting vertical epicyclic motions. 

\section{Velocity Distributions after Orbital Integration}

Using \texttt{galpy}'s Dormand-Prince integrator, the initial particle distributions (with mean velocities shown in Figure \ref{fig:subpini}) are then integrated to the present time (2.2 Gyr for the E2 integration and 1.1 Gyr for the E1 integration). In Figures \ref{fig:subpE1} and \ref{fig:subpE2} we show mean velocities, similar to those presented in Figure \ref{fig:subpini}, 
as well as the particle density distributions as a function of $X,Y$ for both E1, and E2 encounters, but after orbit integration to the present time. The density distributions shown in Figures \ref{fig:subpE1} and \ref{fig:subpE2} resemble those illustrated by previous works using particle integrations \citep{quillen09} and N-body simulations \citep{purcell11,gomez13,widrow14}. We confirm that trailing spiral structure or overdensities can be induced by dwarf galaxy encounters, if the dwarf galaxy mass is similar to $10^{10}M_\odot$. Because self-gravity is not present in our simulations, the spiral structure is due to the radial epicyclic motions excited by the encounters that are evident in the initial perturbations to $V_R$ and $V_\theta$ shown in Figure \ref{fig:subpini}.

In Figures \ref{fig:subpE1} and \ref{fig:subpE2} we also show the mean $Z$ value as a function of $X,Y$.  This is computed in  0.25 kpc square bins in the same way that we compute the mean velocity components. The $\langle Z \rangle$ subpanels in Figures \ref{fig:subpE1} and \ref{fig:subpE2} show that there are large regions of the disc with mean height above or below the galactic midplane, implying that the galaxy disc has become warped. This phenomenon has also been seen in previous simulations (e.g., see Figure 5 by \citealt{gomez13}). The warp arises as vertical amplitudes induced by the encounter (and seen in the $\langle V_Z \rangle$ distributions just after the encounters in Figure \ref{fig:subpini}) have been sheared due to differential precession and the radial gradient of the vertical epicyclic frequency $\nu$. We note that this warp is not self-consistent as we use a static galactic potential, whereas a warp in the disc would influence the potential. The E1 encounter seen in Figure \ref{fig:subpE1} shows ring -- like structure in $\left<Z\right>$ that is found in similar regions as and resembles the Monoceros and Triangulum -- Andromeda overdensities as observed by \citet{xu15} and
\citet{pricewhelan15}. We note that the warp for E2 encounter appears to be more twisted than that of the E1 integration. The E2 encounter occurs 2.2 Gyr ago and there are more rotation periods for the phase wrapping to take place.  

\subsection{Breathing and Bending Coefficients} 

Breathing and bending modes arise from the self-gravity of the Galactic disc. Breathing modes signify that stars above and below the 
midplane move in opposing vertical directions, while bending modes imply stars above and below the midplane move en masse
upwards or downwards \citep{sun15}. 
To quantify bending and breathing modes \citet{widrow14} fit a function to the mean vertical velocity as a function of $Z$ in each $X,Y$ bin,
\begin{equation}
\langle V_Z \rangle (x,y,z) = A_Z (x,y)z + B_Z (x,y) \label{eqn:AzBz}
\end{equation}
(their equation 24). Here $B_Z (x,y)$ is not equivalent to our computed $\langle V_Z \rangle$ as it is equal to the mean value of $V_Z$ at $Z=0$,
and the density distribution may not be symmetric about $Z=0$. 
We use a linear regression to fit for the $A_Z$ coefficient in each 0.25 kpc square bin in $X,Y$. \citet{widrow14} use $A_Z$ to describe breathing modes and $B_Z$ to describe bending modes. The $A_Z$ and $B_Z$ coefficients  are also shown in subpanels of Figures \ref{fig:subpE1} and \ref{fig:subpE2}. 
We have measured $A_Z$ and $B_Z$ coefficients with sizes of several \kmsp \ and \kms \ respectively for both encounters. Figure \ref{fig:subpE1} shows that the structure of the $A_Z \ \textrm{and} \ B_Z$ coefficients in our E1 encounter resembles the spiral structure seen in Figures 10 and 12 in \citet{widrow14}, though it is necessary to note that these figures from \citet{widrow14} occur at different times from our encounters after the initial perturbation. As we have carried out non-interacting test-particle integrations in a fixed potential, these coefficient values and structure cannot be due to bending or breathing waves or modes. Rather, they can only be due to phase wrapping of epicyclic amplitude perturbations.

The perturbing dwarf galaxy is more massive for the E2 encounter, however there is longer time
for phase wrapping to occur and the perturbed motions vary in phase.  
Values for the $A_Z$ coefficient in Figure \ref{fig:subpE2} range from -5 to 5 \kmsp \ but much of
the substructure is erased because the structure is so tightly wound.
For the E1 simulation, structure in $A_Z$ is more coherent and ranges from -10 to 10 \kmsp.
The extent of variations seen in $A_Z$ in our integrations is nearly the same size as those
observed locally; \citet{carlin13} and \citet{sun15} measure $A_Z$ values in the range -10 to 10 \kmsp \ 
(see Figure 13 by \citealt{sun15} and Figure 3 by \citealt{carlin13}). 
Variations in the value of $\langle V_Z \rangle$ are approximately $\pm 10$ \kms \ for both simulations
and similar to those observed (see Figure 14 by Williams et al. 2013).
While phase wrapping from a relatively massive dwarf galaxy with a single encounter
can cause measurable values for the $A_Z$ coefficient, it
may not account for the full extent of the observed values.

\subsection{The vertical gradient of the mean radial velocity component}

\begin{figure*}
\includegraphics[width=6.0in]{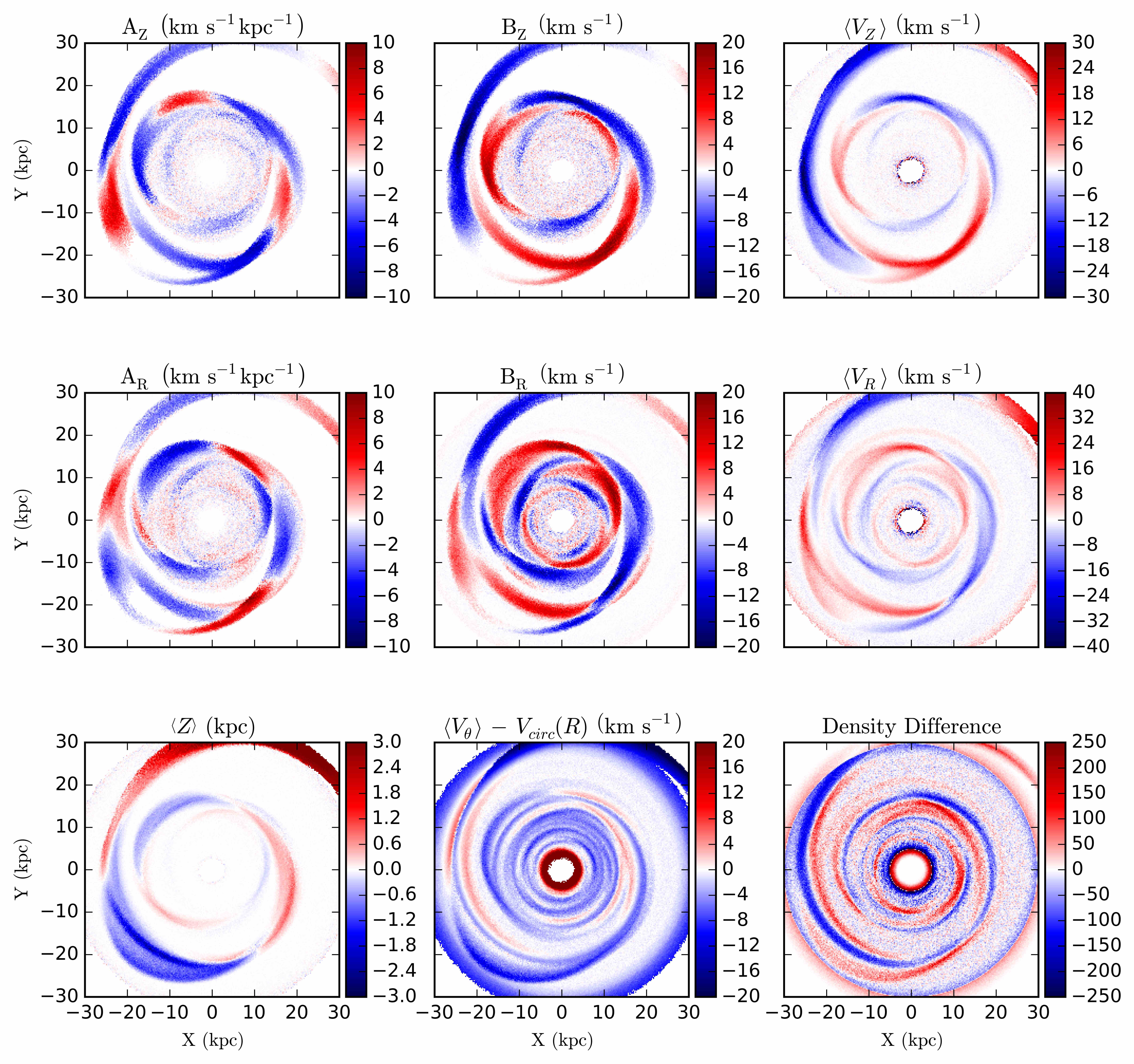}
\caption{
Distribution of mean velocities and $A_Z, B_Z, A_R, B_R$ parameters as a function of $X,Y$ for the E1 encounter
after orbit integration to  the present time. 
Top row from left to right, $A_Z, B_Z$ and $ \langle V_Z \rangle$,
middle row from left to right $A_R, B_R$ and $ \langle V_R \rangle$,
bottom row, $\langle Z \rangle$, $\langle V_\theta \rangle - V_{circ}(R)$ and difference between 
current and initial number densities. 
Velocities and $B_Z, B_R$ are in \kms, and the $A_Z, A_R$ coefficients are in \kmsp.
The difference between current and initial densities is shown in numbers of particles per 0.25 kpc square bin.
\label{fig:subpE1}
}
\end{figure*}

\begin{figure*}
\includegraphics[width=6.0in]{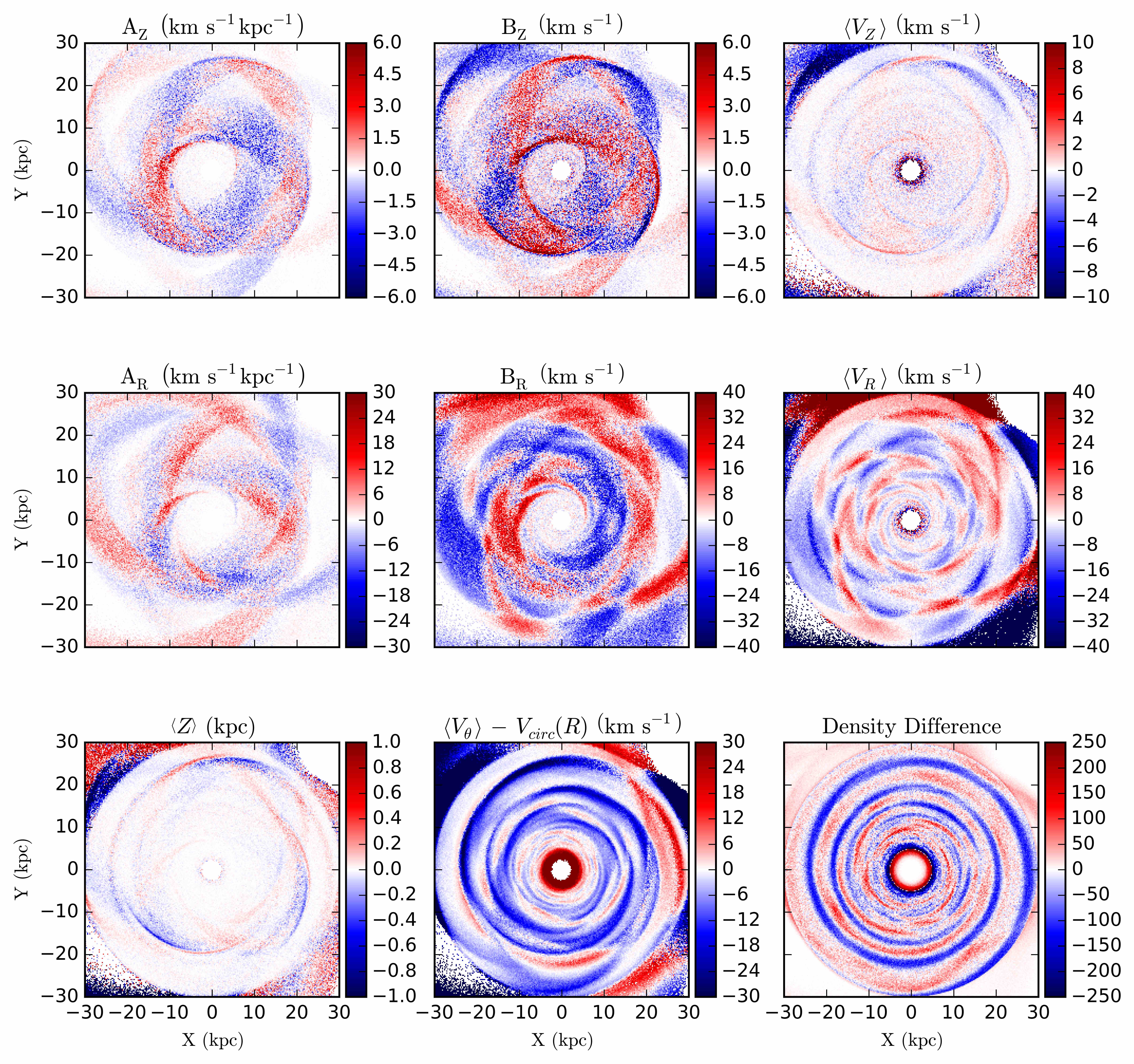}
\caption{
Similar to Figure \ref{fig:subpE1} except for the E2 encounter.
\label{fig:subpE2}
}
\end{figure*}

To quantify the sensitivity of the radial velocities with height above the Galactic plane we also fit a  linear
function to the mean radial velocities 
$$ \langle V_R \rangle (x,y,z) = A_R(x,y)z + B_R(x,y)$$
similar to equation \ref{eqn:AzBz} but using $V_R$ and with coefficients $A_R, B_R$.
These maps are computed in the same way as for the $A_Z, B_Z$ coefficients and also shown in Figures \ref{fig:subpE1}
and \ref{fig:subpE2}.
We can see from these Figures that there are strong correlations between the maps.   

\subsection{Dynamical Interpretation}
\label{sec:dynamics}

Our original velocity perturbations (shown in the initial velocity perturbations in 
Figure \ref{fig:subpini}) are localised near a particular azimuthal angle (near closest approach) 
that we can denote $\Theta_p$, however, $V_R$ and $V_Z$
velocity perturbations also extend over a range of radius.
We can roughly group stars into two sets: those that are perturbed, and those that remain
in circular planar orbits. 
After 1 or 2 Gyr (depending on the encounter) the dependence of the angular rotation rate
on radius, $\Omega(R)$ shears the distribution of perturbed stars.
Meanwhile the stars oscillate vertically and radially, at frequencies that depend on their mean radius or angular momentum.

For an initial perturbation at $\Theta_p$, after a time $\Delta T$
the perturbation will be wrapped in $2 \pi $ in azimuthal angle  across a radial distance 
\begin{equation}
\Delta R \sim \frac{2 \pi}{\Omega_{,r} \Delta T} \sim R \frac{P}{\Delta T}
\end{equation}
where $\Omega_{,r}$ is the radial derivative of the angular rotation rate and 
$P \sim 0.23 $ Gyr is the rotation period at the Sun's Galactocentric radius and we have approximated the rotation
curve as a flat one with $V_c(R) = $ constant. 
For our perturbation with
$\Delta T = 1$ Gyr ago, $\Delta R \sim 1.8 $ kpc.  That means that every 1.8 kpc in radius we should encounter
stars that were initially in the peak of the perturbed region, however this neglects radial motions in the perturbed stars.

\begin{figure*}
\includegraphics[width=6.0in]{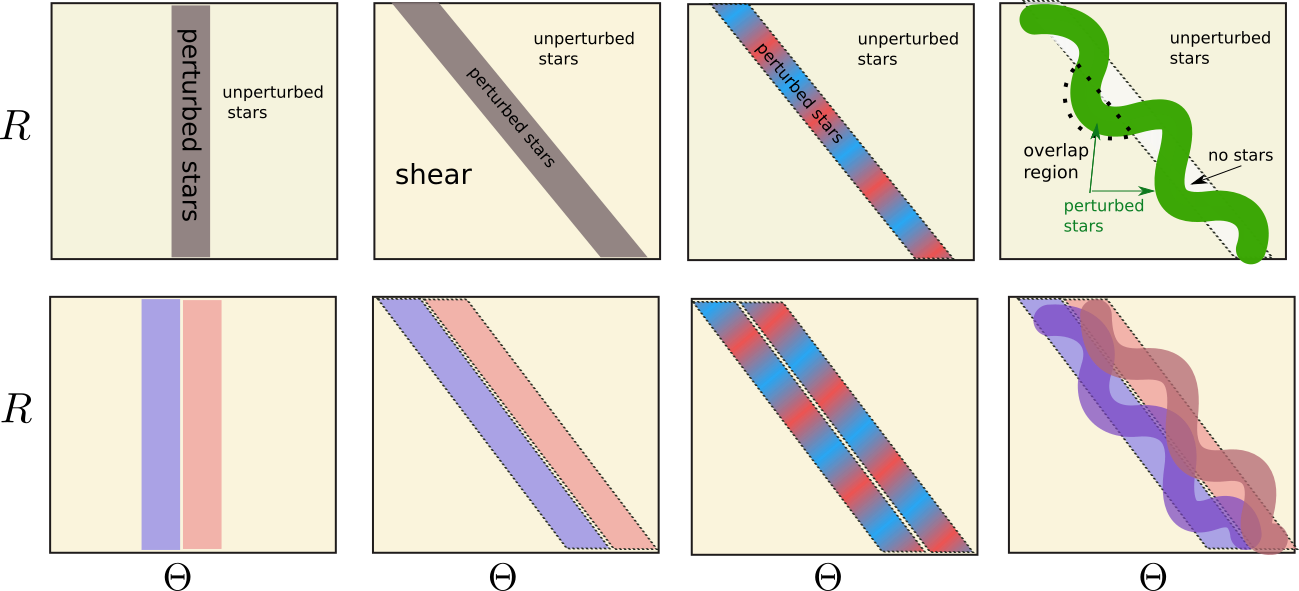} 
\caption{A perturbation localised in $\Theta$ but extending in $R$ is shown on the top left panel.  After a time $\Delta T$
the perturbation shears due to differential rotation (top panel second from left). Taking into account the frequency
of vertical oscillations, the perturbed stars move up and down (top panel second from right).
Taking into account a perturbation in radial
epicyclic and assuming the perturbation has a single phase in radial epicyclic angle $\phi_r$, the radial motions
deform the distribution of perturbed stars giving the green curve on the top right panel.  Where the green curve
lies outside the white bar, perturbed stars are found at the same $R,\Theta$ as unperturbed stars and the combined
populations can give vertical gradients in the velocity components.
The E1 encounter initially has both positive and negative changes to $V_\theta$ giving epicyclic phases that are both positive
and negative (shown on the bottom left panel) and after shearing due to differential rotation (shown in the bottom panel second from left).
Vertical oscillations are in phase and shown in bottom panel second from right.
The radial excursions from each bar have opposite sign and so can overlap one another as well as unperturbed stars
(shown on bottom right panel).
\label{fig:wiggle}
}
\end{figure*} 

The initial perturbation region, in cylindrical coordinates, is illustrated in Figure \ref{fig:wiggle} in the top left panel as a grey bar.
Ignoring the radial motions of the perturbed stars,
after a time $\Delta T$ later,  the distribution of perturbed stars is tilted, as shown in the top  panel second from left.
Stars outside the perturbed region have little radial motion, so they  remain in the region outside the tilted bar but they shear
at the same speed.
For the E1 encounter, initial $V_Z>0$, giving the perturbed stars an initial epicyclic phase $\phi_z \approx 0$.
The grey bar shown in Figure \ref{fig:wiggle} will oscillate vertically as it tilts, after a time $\Delta T$ the perturbed stars
are at different heights (shown in top panel second from right).
Peak to peak vertical maxima should occur across a radial distance
\begin{equation}
\Delta R_\nu = R \frac{P}{\Delta T} \frac{\Omega}{\nu}.
\end{equation}
and computing this distance for 
$\Delta T = 1$ Gyr and $\nu/\Omega \sim 2.6$ (using \texttt{galpy}'s Milky Way potential \texttt{MWPotential2014 })
we estimate $\Delta R_\nu \sim 0.75$ kpc. This we expect is the radial distance
between two maxima in $Z$ in the inner region of the galaxy, and it is approximately consistent with
what is seen near $R_\odot$ in the lower left panel of Figure \ref{fig:subpE1} showing $\langle Z \rangle$.
We compare the peak to peak radial distance from epicyclic phase wrapping
to the wavelength predicted for bending waves, $\Delta R_\nu \sim 10$ kpc (\citealt{weinberg91}, see Section 6).
The radial distance between peaks in $Z$ can be much smaller than the wavelength predicted for bending waves.

The perturbed stars also oscillate radially.  If there is a single
phase for $\phi_r$ for the perturbed stars, then the distribution of the perturbed stars wiggles, as shown
by the green curve on the top right panel. The wiggles in the green curve can lie in the same regions as stars on
circular orbits (yellow background). There are overlap regions that would contain higher surface densities of stars
accounting for the similar morphology of the density panel and velocity panels in Figure \ref{fig:subpE1}.  
For some $R,\Theta$ there are two populations of stars:
stars in circular orbits and stars that were perturbed by the encounter. As the stars perturbed
by the encounter can reach larger distances above and below the Galactic plane, the combined population can
exhibit a vertical velocity gradient. 
Likewise, there are regions where perturbed stars outnumber unperturbed stars, leading to underdensities that can 
still exhibit velocity gradients. 

The E1 encounter also gives velocity perturbations in $V_R$ and $V_\theta$, resulting in a region of
the galaxy with stars that all have an epicyclic
angle $\phi_r \sim -\pi/2$ for $\Theta$ slightly lower than the center $\Theta_p$ and $\phi_r \sim \pi/2$ for $\Theta$ slightly higher than the center,
the change in sign resulting from the azimuthal variation in $V_\theta$ (see Figure \ref{fig:subpini}).
 We can consider a localised (in $\Theta$) increase in 
epicyclic amplitudes $a_r$ and $a_z$, with $\phi_z=0$, near the center and $\phi_r = -\pi/2$ but flipping sign
at $\Theta_p$. This situation is represented in the lower left panel in Figure \ref{fig:wiggle} and after shearing in $\Theta$
in the lower panel, second from left. Vertical oscillations are in phase and shown in the lower panel second from right.
Because 
the epicyclic frequency, ($\kappa$) and vertical oscillation frequency ($\nu$) depend on radius  
as the region of perturbation is stretched, it also oscillates radially and up and down. 
A perturbation in the radial direction and initially all in phase in $\phi_r$
will be wrapped in its radial epicyclic by $2 \pi$ (corresponding to two particles each at pericentre) 
across a radial distance 
\begin{equation}
\Delta R_\kappa = R \frac{P}{\Delta T} \frac{\Omega}{\kappa}
\end{equation}
Using $\kappa/\Omega \sim 1.4$ we estimate  $\Delta R_\kappa \sim 1.3$ kpc near the Sun.
If all perturbed particles were initially in phase then we would expect the distance between radial oscillation maxima
to be 1.3 kpc, however the $\phi_r$ varies by almost $\pi$ in the initial perturbation and this gives
about twice as many oscillations.  Furthermore, perturbed stars can overlay each other as well as overlay
unperturbed stars (as shown in bottom right panel in Figure \ref{fig:wiggle}).
The two effects likely account for the increased number of red and blue regions in the $A_R$ and $B_R$ panels compared to 
 the $A_Z$ and $B_Z$ panels shown in Figure \ref{fig:subpE1}.

 In Figure \ref{fig:subpE1} the $\langle V_Z \rangle$ and $\langle Z \rangle$ subpanels
resemble each other except they are $90^\circ$ out of phase, as would be expected from vertical oscillations of the perturbed
stars.  The $A_Z$ slope coefficient measures the gradient of $\langle V_Z\rangle $  with $Z$.
In regions where stars in the midplane coexist with (or overlap)  perturbed stars, a gradient is  measured
 when the perturbed stars lie above or below the Galactic plane and there are unperturbed stars in the midplane.
 Where the perturbed particles are above the plane ($Z>0$) but moving downward ($V_Z <0$), we estimate the gradient 
 from these stars and the unperturbed ones in the midplane with $V_Z\sim 0$, giving approximately 
 $A_Z \sim \langle V_Z \rangle /\langle Z \rangle$ and
 $A_Z$ is negative. For stars below the plane ($Z<0$) with $V_Z<0$, $A_Z>0$. Thus, we expect the $A_Z$ map
 contains twice as many color features as the $\langle Z \rangle$ and $\langle V_Z\rangle $ maps
 and indeed that is what we see on the top left in the $A_Z$ panel of Figure \ref{fig:subpE1}, when compared to 
 the $\langle V_Z \rangle$ and $\langle Z \rangle$ panels.  
 
 Similar phenomena are seen in Figure \ref{fig:subpE2}, showing the E2 encounter, except there are 
 many more features in each panel.
 This is expected as the time to wrap in phase is twice as long, but in addition the initial perturbation
 contains regions differing in phase $\phi_z$ and $\phi_r$ also contains additional structure
 as initial $V_R$ and $V_Z$ distributions are more complex than for the E1 encounter (see Figure \ref{fig:subpini}).
 Because the E2 encounter is less effective at inducing vertical oscillations and much of induced structure
 is erased due to the long timescale and higher complexity of the epicyclic phases, hereafter we primarily
 discuss the E1 encounter.
 
Noticeable correlations between the density of stars and different velocity components exist in our simulations. 
Figures \ref{fig:subpE1} and \ref{fig:subpE2} show that regions that have extreme velocity means are the same
as regions with extreme gradients, and these are the same as those with high or low surface densities. 
Perturbed stars lie chiefly in regions where the mean velocity component is nonzero and there are overdensities or underdensities. 
As perturbed stars lack circular orbits, they have radial velocity 
components that affect their tangential velocities; in general, positive $\left<V_R\right>$ gives negative 
$\left<V_\theta \right> - V_{circ}(R)$, and this occurs where there are overdensities in our integrations. 
Differences in bulk radial velocities can result in separated overdensity streams, where bands of stars share the 
same density but have split off from each other, such as those seen around 
$(X, Y) = (-15, 15)$ kpc and $(X, Y) = (5, 10)$ kpc in the density panel in Figure \ref{fig:subpE1}. 
These disparities in $\left< V_R \right>$ also produce bifurcations, or `forked - tongue' patterns, in $\left<V_Z\right>$,
seen in the upper right panel in Figure \ref{fig:subpE1}. 

Negative values of $\left<V_\theta \right> - V_{circ}(R)$ (corresponding to positive asymmetric drift) 
are expected from a steady state disk, though
here  the  density, mean velocity components and dispersions are  time dependent.
Because the initial density distribution is proportional to $1/R$ we expect that eccentric particles
are more likely to affect the velocity mean near apocenter where $V_\theta $ is lower than
that of particles in circular orbits, rather than near pericenter where $V_\theta$ is higher than 
that of particles in circular orbits.

Vertical velocity gradients are seen in our numerical integrations and they can be explained
by vertical and radial epicycle motions  excited by the encounters.
The vertical gradients  arise in regions where perturbed and unperturbed stars (or different
source regions of perturbed populations)
lie in the same $X,Y$ region.  Because they arise due to overlap of different populations we expect
(and see in our integrations) strong correlations between the distributions of the different velocity
components and the density. The number of peaks in $\langle Z \rangle $ or $\langle V_Z \rangle $ can roughly
be estimated using a shearing timescale based on the oscillation frequency
and the timescale since the perturbing encounter.

\subsection{Local Velocity Gradients}

\begin{figure*}
\centering
\includegraphics[width=6in]{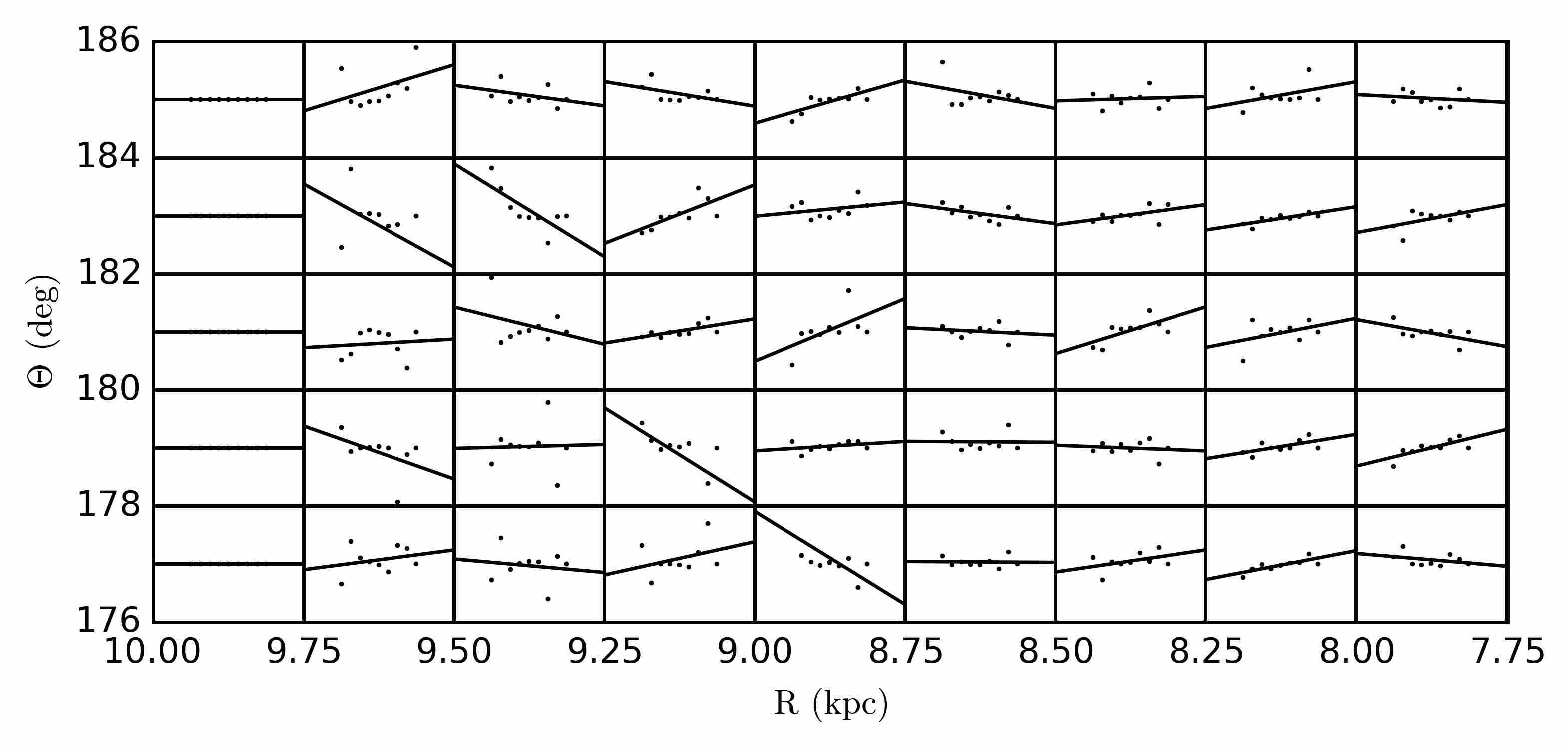}
\caption{Mock observations of the solar neighbourhood for breathing and bending mode parameters $A_Z \ \textrm{and} \ B_Z$ similar to \citet{sun15} Figures 13 and 14 for encounter E1.  
Each subpanel is at a different central $\Theta$ and $R$ with values for $\Theta$ and $R$ shown on the left and bottom.
The $X$ axis of each subpanel has range $-2 \leq Z \leq 2$ kpc and the $Y$ axis $-30 \leq \langle V_Z\rangle \leq 30$ \kms, the linear fits showing vertical gradients in $V_Z$. This local neighbourhood shows smooth variations in mean velocity components
as a function of $Z$. The gradients in the mean velocities vary rapidly with position in the simulated disc. 
\label{fig:ps_sun_lines}
}
\end{figure*}

To mimic observations of nearby stars in the Galaxy we can extract test particles in small regions of
 the Galaxy. \citet{sun15} (in their Figures 13 and 14) showed linear fits to $\langle V_Z \rangle $ with $Z$ at different locations in $X,Y$ near the Sun.    
We construct here a similar figure to illustrate that our
integrations display gradients in the vertical velocity component.
A  matrix containing numbers of particles is constructed in $\left(8 \leq R \leq 10 \ \textrm{kpc}, 176\degree \leq \Theta \leq 184\degree, -1 \leq Z \leq 1 \ \rm{kpc}\right)$ with binning size $\left(\Delta R, \Delta \Theta, \Delta Z\right) = (0.25 \ \rm{kpc}, 2\degree, 0.25 \ \rm{kpc})$. After generating the number density matrix, we apply the same outlier rejection criteria used by \citet{sun15}, namely requiring $-200 < V_R < 200$ \kms, $0 < V_\theta < 400$ \kms \ and $-200 < V_Z < 200$ \kms \ to compute a velocity matrix in $\langle V_Z \rangle$; we then divide the velocity matrix by the number density matrix to yield a bulk or mean $V_Z$ matrix in $R, \Theta\ \textrm{and} \ Z$. We use linear regression to find the best fit line of 
$$\left<V_Z\right> (R, \theta, z) = A_Z (R, \theta) z + B_Z (R, \theta) $$ 
for each pair of $\left(R, \Theta \right)$. The individual values for $\langle V_Z \rangle (z)$ are shown in Figure \ref{fig:ps_sun_lines} at $\Theta = 180^\circ$ along with the linear fits giving us $A_Z$ and $B_Z$ coefficients; the $X,Y$ axes of each subpanel correspond to $-2 \leq Z \leq 2$ kpc and $-30 \leq \left<V_Z \right> \leq 30$ \kms, respectively. In the previous section we attributed the
velocity gradients to overlap of populations of perturbed and unperturbed stars. However, most of the panels in Figure 
\ref{fig:ps_sun_lines} show smooth variations in mean velocities as a function of height above and below the Galactic midplane.
Had we started with a planar thin disc in circular orbits we would have seen folds in the disc and these would not have given 
smooth velocity gradients. We can attribute the smooth slopes to the velocity dispersion in our initial disc. 
\citet{sun15} (and others) measured different velocity gradients in different regions of the galaxy. Figure \ref{fig:ps_sun_lines} 
shows that there are rapid variations in velocity gradients with position in our simulated Galaxy. Hence some or all 
of the fine structure measured by \citet{sun15} and \citet{carlin13} might be real, and we might expect that future
observations extending the number of stars, precision and distance of stars will uncover even more structure in the velocity field. 

 
We now explore whether our integration can exhibit the patterns seen as a function of $Z$ in recent observational studies.
 A key feature seen in the local velocity distributions is 
 a strong vertical gradient in $\langle V_Z \rangle$  with negative $V_Z$ at positive
$Z$ and a slope of  $A_Z \sim -5 $ to $-10$ \kmsp \    (Carlin et al. 2013; Williams et al. 2013). 
Regretfully,
in Figure \ref{fig:subpE1} $A_Z >0$ at $(-R_\odot,0,0)$ and so is not consistent with the observed pattern of gradients near the Sun.
As the gradients are comprised of tightly wound structures, they are sensitive to the time since the encounter.  We can 
effectively vary the time since encounter by considering different angular locations in the Galaxy at the same radius as the Sun.
The rotation period at the Sun is $P \sim 0.23$ Gyr, hence an uncertainty in the time since the E1 encounter of only 0.1 Gyr
would give rotation by nearly $180^\circ$. Estimates of the time since the last pericentre encounter range from 0.8 to 1.1 Gyr. 
For example, \citet{purcell11} give a time and distance for last pericentre for the Sagittarius dwarf galaxy of 
$t_{peri} \approx 0.85 \ \textrm{Gyr ago}, R_{peri} \approx 15$ kpc for their light model and 
$t_{peri} \approx 1.1 \ \textrm{Gyr ago}, R_{peri} \approx 15$ kpc for their heavy model. \citet{law10} in their Figure 7
give $t_{peri} \approx 0.8 \ \textrm{Gyr ago}, R_{peri} \approx 15$ kpc for the Sagittarius dwarf. 

\begin{figure*}
\centering
	\includegraphics[width=5in]{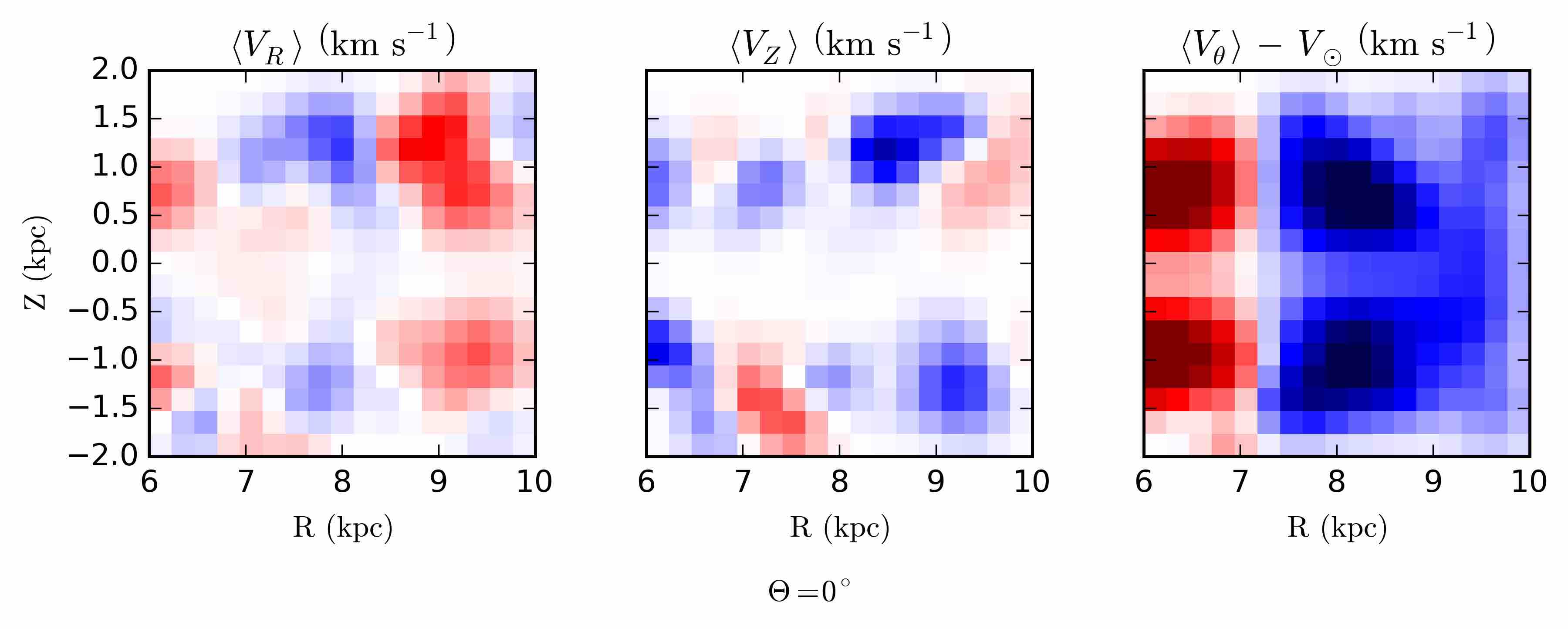}
	\includegraphics[width=5in]{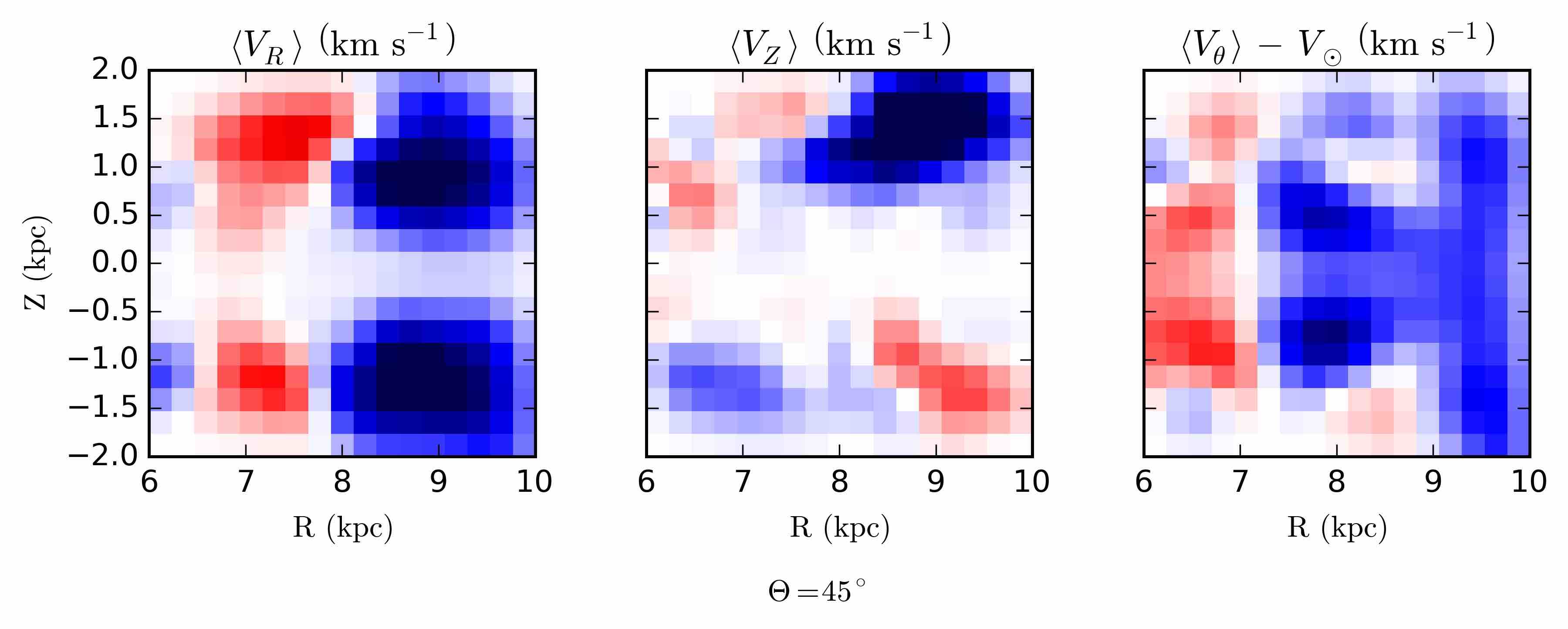}
	\includegraphics[width=5in]{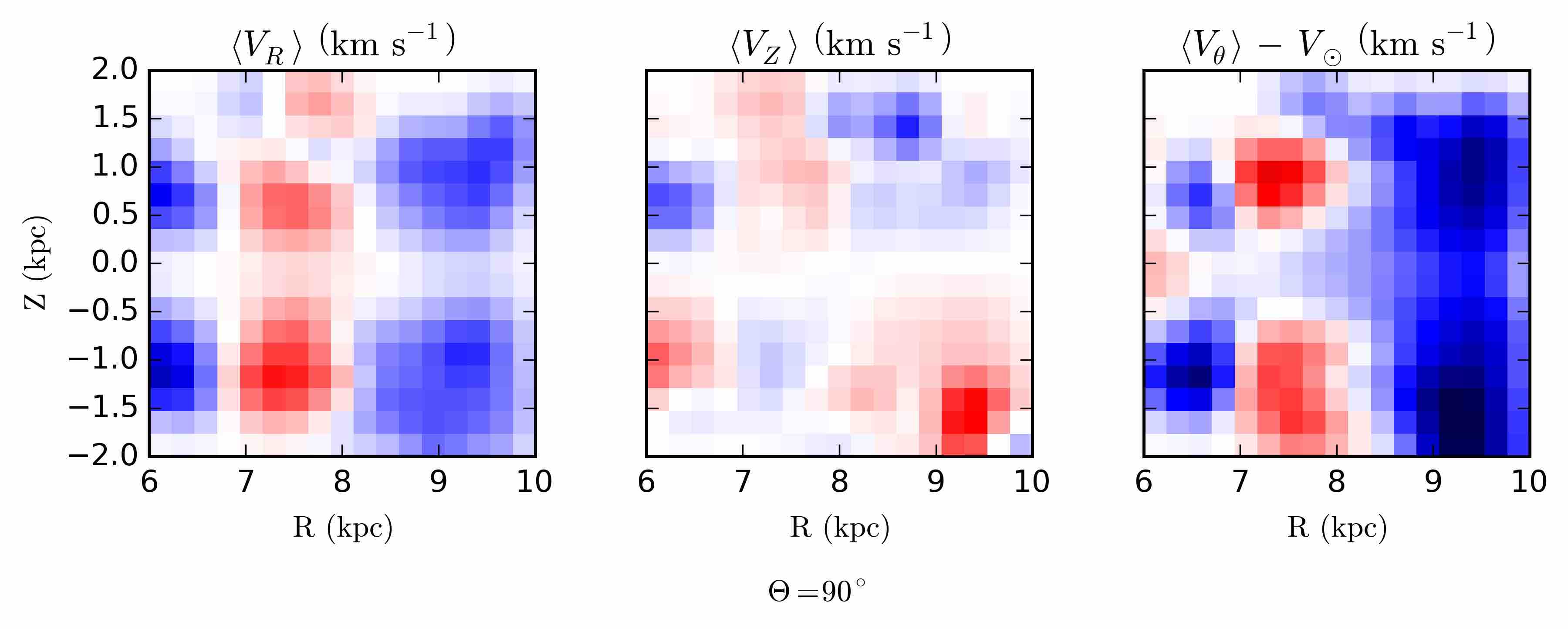}
	\includegraphics[width=5in]{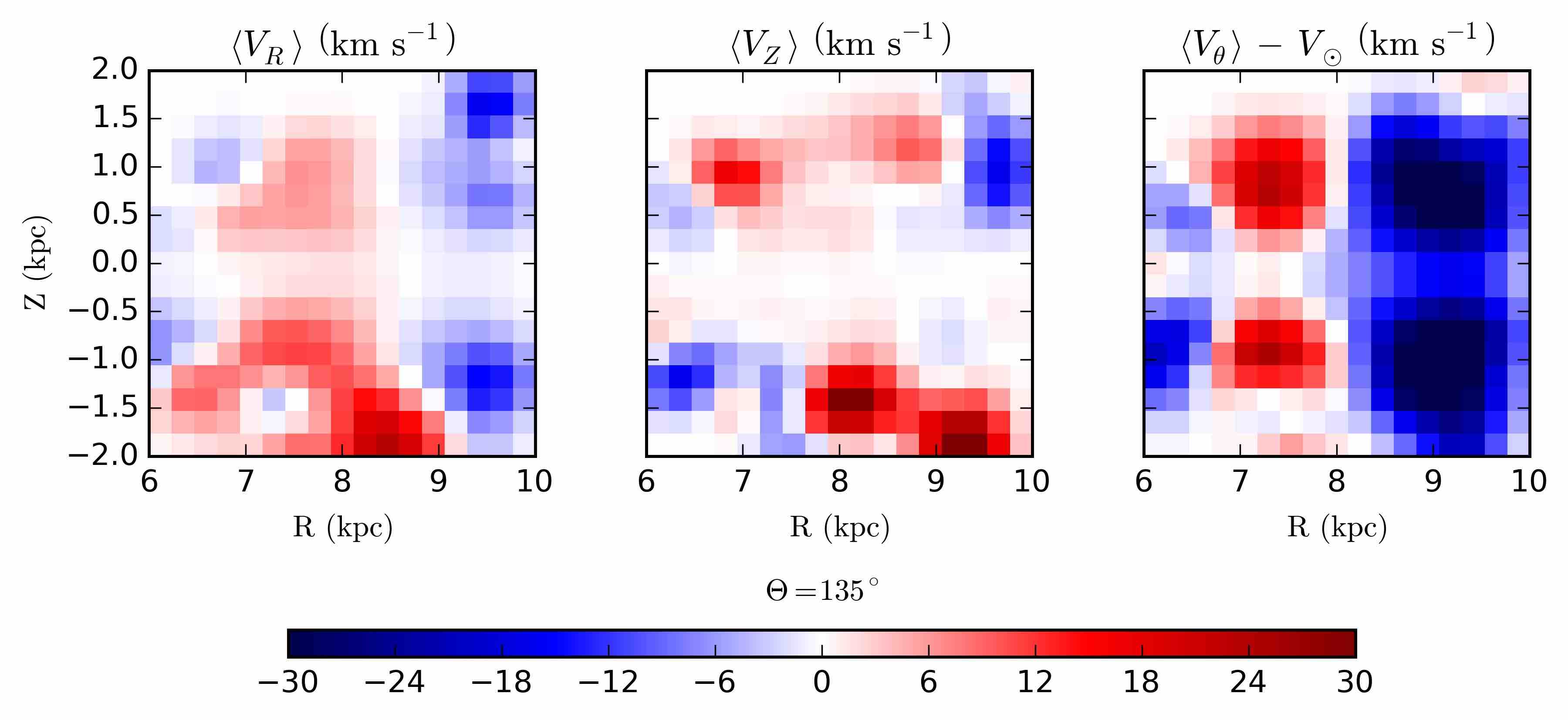}
\caption{Mock observations of mean velocity components in local regions as a function of $R$ and $Z$  for simulation E1.
Each row shows (from left to right) $\langle V_R \rangle$, $\langle V_Z \rangle$, and $\langle V_\theta \rangle - V_{circ}(R) $.
Mean velocities have been extracted at different 
azimuth angles. Panels from top to bottom have central angle $\Theta = 0, 45, 90 \ \& \ 135$\degree, respectively. 
The $\left<V_Z\right>$ panel at $\Theta = 45^\circ$ displays a particularly large gradient. 
\label{fig:ps_will_0}
}
\end{figure*}

In Figures \ref{fig:ps_will_0}, we emulate Figures 10, 12 and 13 by Williams et al. (2013) and Figure 2 by \citet{carlin13} by plotting gradients of $\left<V_R \right>, \left<V_\theta \right> - V_{\odot}, \textrm{and} \ \left<V_Z \right>$ in different local neighbourhoods for encounter E1 in $(R, Z)$ with $6 \leq R \leq 10, -2 \leq Z \leq 2$ kpc and $\Theta$ in 45\degree \ increments with a 16\degree \ spread using bin sizes of $(\Delta R, \Delta Z) = (0.25, 0.25)$ kpc. We generate mean velocity matrices in the same method as previously above.  The resulting mean velocities are smoothed with a Gaussian filter with $\sigma_Y, \sigma_X = 1$ to reduce the graininess resulting from the particle distributions. 
 Like Figures \ref{fig:subpE1} and \ref{fig:subpE2} we only show in Figure \ref{fig:ps_will_0} 
the distribution at a particular time since encounter, but the structure would
be different if the encounter occurred earlier or later.
Figure \ref{fig:ps_will_0} shows that there is potentially quite a bit of structure in the bulk velocities above and below the Galactic plane, and that there could be large variations in the vertical gradients as a function of $R, \Theta$ as well as $Z$.
If phase wrapping of epicyclic perturbations is responsible for the observed variations in bulk velocities, a consequence would
be predicted variations in the bulk velocity as a function of distance from the Sun, such as the symmetry in $\left<V_R\right> \ \textrm{and} \ \left<V_\theta \right> - V_\odot$ above and below the midplane. 

Figure \ref{fig:ps_will_0} illustrates that there are regions in our simulated galaxy that show negative 
vertical velocity gradients in $\langle V_Z \rangle$.
For example, the second panel corresponding to $\Theta=45^\circ$, displays negative $V_Z$ above the Galactic plane for $R>8$ kpc
and vice versa below the Galactic plane, giving $A_Z$ with the expected and observed sign and about  the right size.
The velocities near the midplane in most regions are approximately zero, consistent with the presence of unperturbed stars
in the extracted neighbourhoods. The largest velocities are seen away from the Galactic plane, consistent with the
presence of perturbed stars.  
However, as we did not include a thick disc component in our initial particle distributions, above and below the Galactic plane
are found only perturbed particles and these dominate the mean velocities. Had we included a thick
disc component we could have measured lower gradients as they could have diluted the contribution from the perturbed stars.  
We adopted a $1/R$ number density
to increase the number of particles at large radius, and this too would bias our computed values of $A_Z$ and $B_Z$.

\citet{carlin13} show negative $\langle V_R \rangle \sim -5$ \kms \ near the midplane and positive $\langle V_R \rangle \sim 5$ \kms \ 
above and below the midplane. Our integrations primarily show a zero value of $\langle V_R\rangle$ in the midplane
because unperturbed disc particles are in nearly circular motion. However, in many of the rows 
shown in Figure \ref{fig:ps_will_0} the $\langle V_R \rangle$ distributions
are symmetrical at positive and negative $Z$, such as in the plot at $45^\circ$. We attribute this to the fact that $\nu \sim 2 \kappa$
throughout the disc, so vertical epicyclic motions have sheared out more than have radial epicyclic motions, leaving the radial structure more coherent. 
This symmetry is exhibited in numerous extracted regions in Figure \ref{fig:ps_will_0},  though the $45^\circ$ panel
shifts between negative and positive $V_R$ at about $R=8$ kpc.
\citet{carlin13} shows a negative value for $V_R$ at $|Z|<300$ pc so the velocities could be associated with
self-gravitating spiral density waves, and these are not present in our integrations.
One last thing to point out is that the $\langle V_\theta \rangle - V_\odot$
panels show regions that are both positive and negative above and below the plane (usually symmetrical above and below
the plane).
Asymmetric drift associated with a thick disc high velocity dispersion would lower
the tangential velocity component. If perturbed populations of stars dominate
at high and low galactic latitude, one would have to take care not to misinterpret
the mean tangential velocity components.

In summary, mock observations of our simulation at $R_\odot$ and different azimuths show that perturbing the disc with a fairly 
massive Sagittarius dwarf yields measurable gradients with similar sizes compared to recent observations. The vertical velocity component
varies smoothly with $Z$f and has slope that varies quickly with position in the Galaxy (Figure \ref{fig:ps_sun_lines}), mimicking
measurements by \citet{sun15}. Mean velocities as a function of $\left<Z\right>$ (Figure \ref{fig:ps_will_0}) display gradients and 
symmetry above and below the midplane similar to those found by \citet{carlin13,williams13}, however we do not identify
a particular region in our simulated galaxy that matches the gradients of all observed velocity components.

\section{Summary and Conclusion}

Non-interacting test particle integrations in a fixed potential are  less accurate than N-body simulations. However, they have the advantage that the role of phase-wrapping of epicyclic perturbations can be studied in isolation from phenomena that require self-gravity such as bending and breathing waves and modes. We start with a thin disc containing stars subjected to a velocity impulse computed from an instantaneous hyperbolic orbit approximation from a single close encounter with a dwarf galaxy. Particles are then integrated forward to the present time in a fixed Galactic potential and during this time the distribution of epicyclic angles is progressively sheared due to the dependence of vertical and radial epicyclic oscillation frequencies on mean orbital radius. To illustrate this process we use two encounters taken from estimated orbits for the Sagittarius dwarf galaxy nucleus. After the orbit integration the mean velocities of the stars display vertical velocity gradients that have been previously interpreted in terms of breathing and bending modes or waves.  
However, because our integrations lack self-gravity, these phenomena must instead be due to shearing or phase wrapping of epicyclic perturbations
caused by the encounters.

Because the encounter excites radial perturbations, the perturbed particle distribution not only shears with azimuthal angle
but wiggles, overlapping the distribution of unperturbed particles.
We attribute the measured vertical gradients in bulk velocities to regions where
perturbed and unperturbed particles overlap or are found at the same projected $X,Y$ in Galactocentric coordinates, 
or  regions where primarily perturbed stars are located, where underdensities would occur.  
A vertical gradient arises when a population of
perturbed stars lies above or below the plane and its velocity components are compared to those of unperturbed particles
that lie in the midplane. Vertical gradients could also arise when different populations of perturbed stars 
overlie the same $X,Y$ location. This scenario predicts correlations between vertical gradients of different
velocity components and density. The number of peaks in radial or vertical velocity means can be estimated
from the time since the encounter and the phases in epicyclic angles induced by the encounter.

This work solely considers epicyclic phase wrapping on the Galaxy after an encounter from an external perturber.
Effects from internal perturbations, such as spiral arms or a bar, or those due to self-gravity, such
as breathing and bending modes, are not present in our integrations.   Because phase wrapping of epicyclic
perturbations can produce vertical gradients in bulk velocities, breathing and bending modes are not required
to account for these gradients.   The wavelength of bending waves near the Sun has been estimated to
be quite large ($\sim 10$ kpc \citealt{weinberg91}) implying that variations in bulk velocities over short distances
must be due to coherent epicyclic motions, rather than self-gravitating waves or modes. 
That is not to say that self-gravity has no effect.   
Internal structures such as spiral density waves or the Galactic bar
might also induce velocity gradients \citep{faure14,monari15}.
An external perturber in the outer disc would produce gradients in bulk velocities that increase in magnitude with increasing distance from
the centre of the galaxy, while an internal perturber (such as the bar) would exhibit gradients with decreasing magnitude with larger radius 
(e.g., \citealt{monari15}), so future observations could differentiate between the two mechanisms. 
The radial distance in maxima of $\left< Z \right>$, compared to bending wavelengths, may make it possible to distinguish
bending waves from epicyclic phenomena. Unfortunately, studies of breathing modes have been restricted to the plane parallel 
setting \citep{widrow14, widrow15} and so do not predict structure as a function of position in the Galaxy. 

We find that the simulated sizes of the $A_Z$ and $B_Z$ coefficients, used to quantify gradients of bulk vertical motions by 
\citet{widrow14}, are similar to those observed by \citet{sun15} for a moderate mass Sagittarius dwarf galaxy, $\sim 2.5 \times 10^{10} 
M_\odot$, that passed its orbital pericentre about a Gyr ago.  The passage of the Sagittarius galaxy through the Galactic plane 
approximately 2 Gyr ago is less effective than the pericentre at approximately 1 Gyr ago for three reasons. The extent of phase 
wrapping is more extreme and warping more tightly wound. Secondly, the E2 perturbation itself excites vertical epicyclic motions that 
vary in phase by $\pi$, whereas the pericentre encounter 1 Gyr ago, because it passes above the disc, excites vertical epicyclic 
motions in phase.  Thirdly, even though we used a larger mass for the E2 encounter, the size of vertical velocity perturbations is 
smaller than that for the E1 encounter because of the orientation of the encounter.  

Regions at a solar neighbourhood radius extracted from our E1 integrations can show negative values of the $A_Z$ coefficient,
similar to that recently measured in the Solar neighbourhood (Carlin et al. 2013; Williams et al. 2013) but not at our expected 
location of the Sun in the simulation.
However, a small error in the time estimated since the latest encounter could account for this discrepancy. 
We fail to find a local region in the E1 integration that matches the observed vertical gradients of all velocity components.
Our integrations do illustrate that phase wrapping of epicyclic perturbations caused by
the Sagittarius dwarf galaxy might account for  much of structure seen
in bulk velocity motions away from the Galactic plane, and if so, there should be large variations in the bulk motions with position
in the Galaxy.

We simulate two perturbations from the Sagittarius dwarf galaxy, each with different mass at different times and positions, from one 
orbit of the dwarf galaxy. If the dwarf is massive enough that dynamical friction is important, then the initial mass of the Sagittarius 
dwarf affects the orbit \citep{purcell11}, further contributing to the uncertainties in orbital parameters for our encounters. It is possible 
that the Galactic disc could have been more recently perturbed by an as yet unidentified dwarf, as proposed by \citet{chakra09}. 

In this paper we have used highly simplified integrations to isolate epicyclic phase wrapping from other phenomena.
The impulse approximation for the encounters could overestimate the energy transfer to the Galactic disc \citep{donghia10}, as disc response 
and time dependence of encounters have not been taken into account. Our orbit integration neglects 
spiral structure and associated radial migration that would vary oscillation frequencies and so cause a loss of 
coherence in the epicyclic phases(e.g., \citealt{veraciro15}).
 Self-gravity would also cause variations in the epicyclic amplitudes and angles and so might need
to be taken into account to relate structure in the velocity field to a previous encounter.
A future comparison between simulations with gravitationally interacting particles and non-interacting particles could
be used to identify and study possible bending waves or breathing modes that might be present in N-body simulations.
Test particle simulations could be used to improve understanding of the role of the time dependence of the encounters
and relate the size scale and distribution of velocity gradients and bulk motions to the perturbations themselves.
Thus, improved numerical models and observations might together match the observed gradients and simultaneously better
constrain the mass and orbit of the Sagittarius dwarf galaxy.

\section*{Acknowledgements}

We thank Ivan Minchev for valuable comments and the University of Wisconsin, Madison for hospitality in spring 2015.
This work acknowledges support from REU program NSF grant NSF-PHY 1460352
and was in part supported by NASA grant NNX13AI27G.
ED gratefully acknowledges the support of the Alfred P. Sloan Foundation and the hospitality
of the Aspen Center for Physics, funded by NSF
Grant No. PHY-1066293. This work is partially funded by NSF
Grant No. AST-1211258 and ATP-NASA Grant No. NNX14AP53G.


\setcounter{figure}{7}
\begin{figure*}
\centering
	\includegraphics[width=5in]{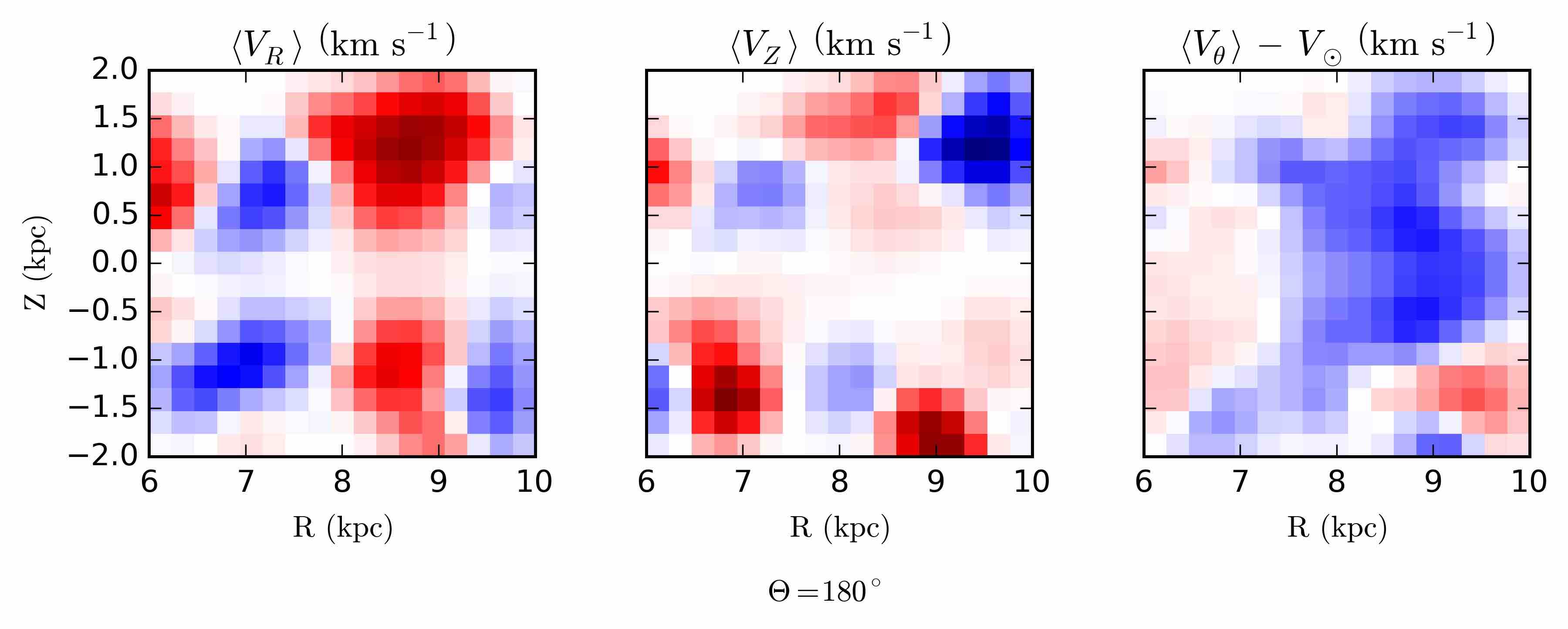}
	\includegraphics[width=5in]{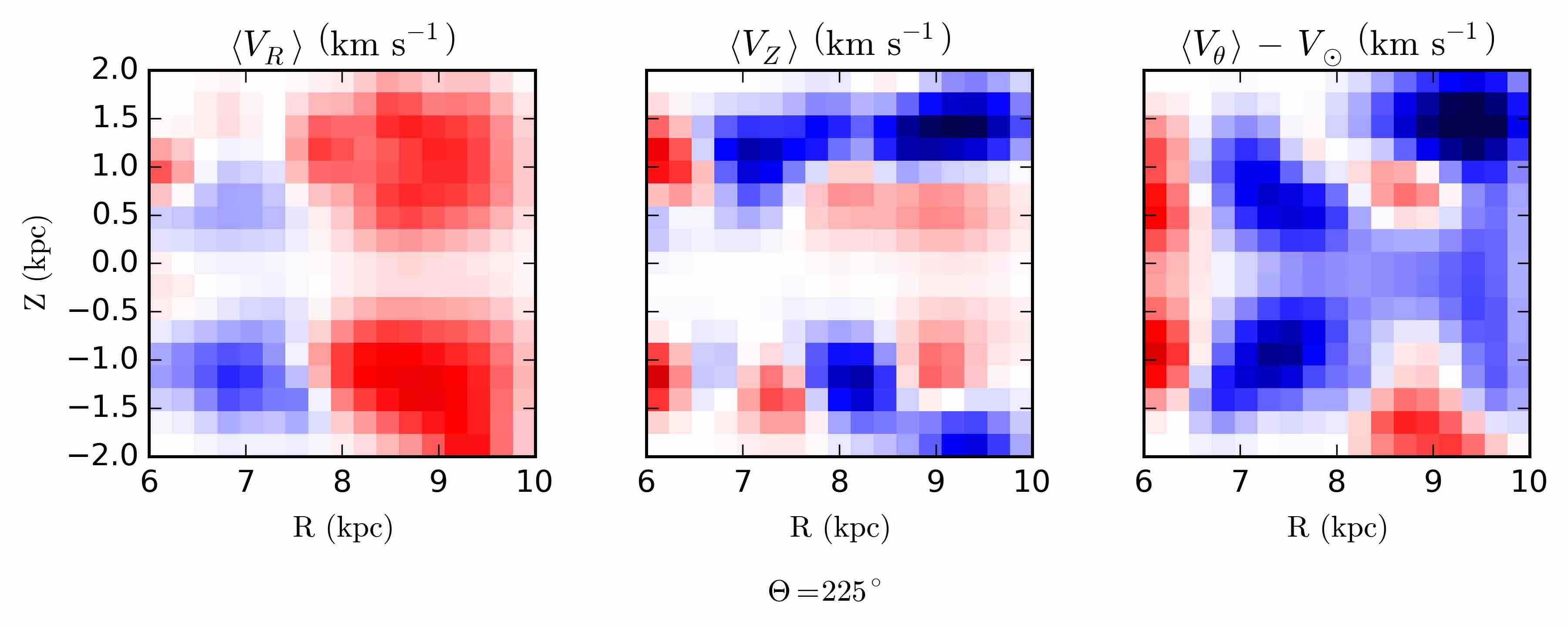}
	\includegraphics[width=5in]{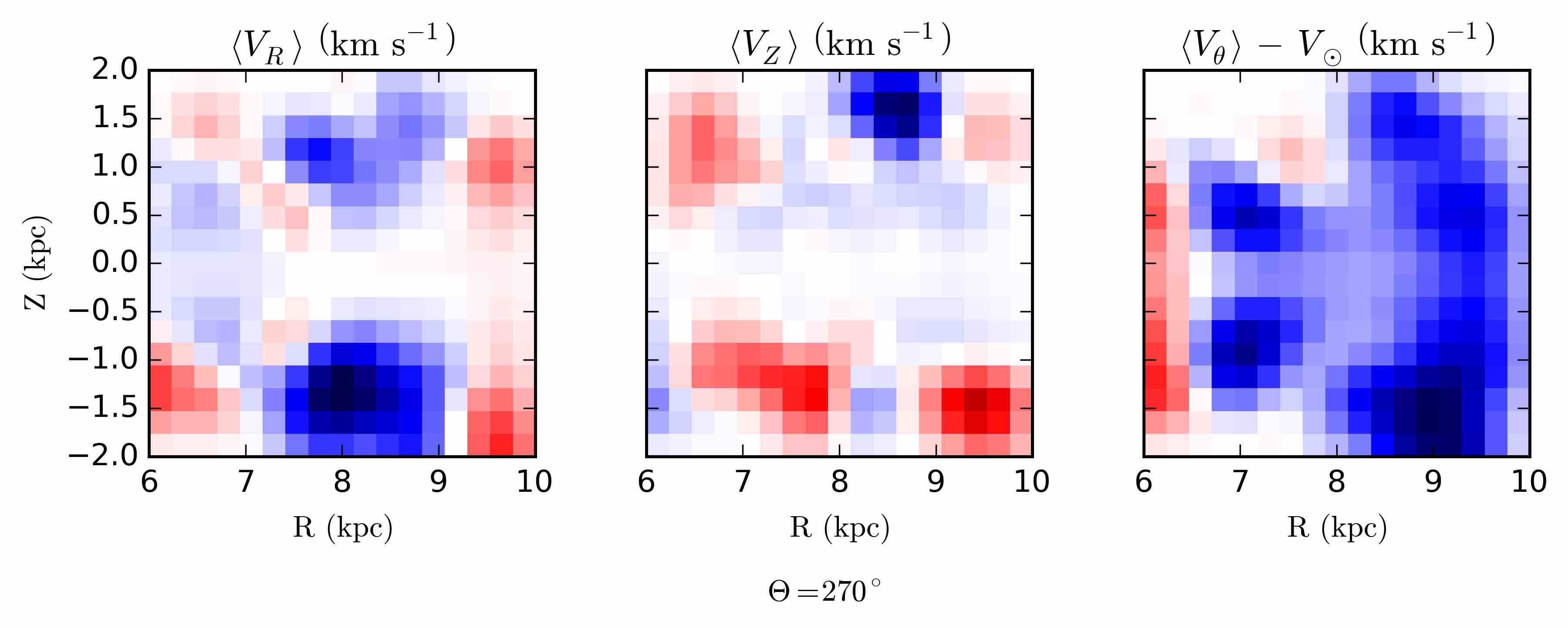}
	\includegraphics[width=5in]{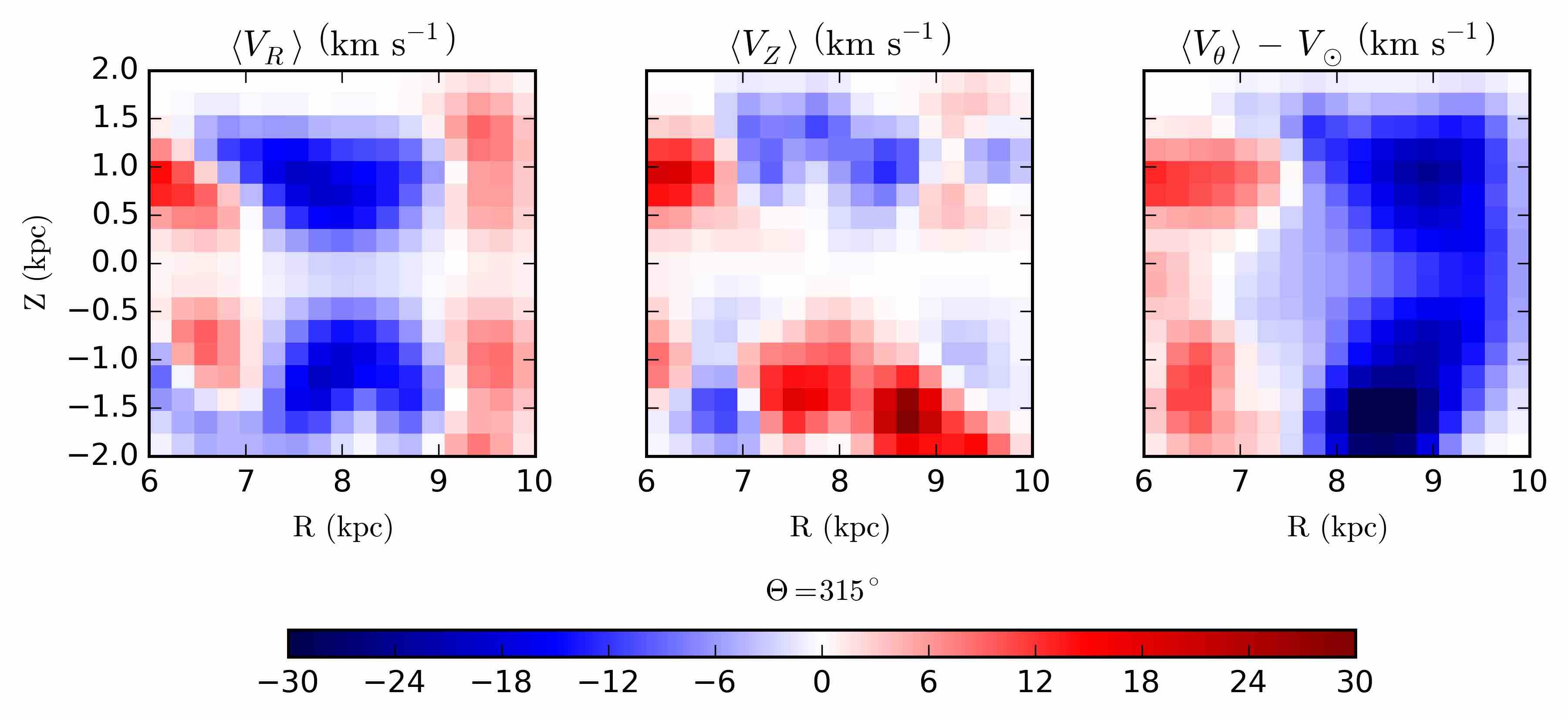}
\caption{Figure \ref{fig:ps_will_0} continued, with  (top to bottom) $\Theta = 180, 225, 270 \ \& \ 315$\degree.
}
\end{figure*}

\appendix
\section{Instantaneous hyperbolic orbit approximation for velocity perturbations}
\label{ap:hyper}

We approximate the perturbation caused by the passage of the dwarf galaxy through the Galaxy
midplane (E2 encounter) and pericentre encounter (E1 encounter) with a velocity impulse.  
Each star's velocity is changed instantaneously. 
The velocity perturbation is estimated assuming that the encounter can be approximated 
with a hyperbolic orbit (e.g., see  (\citealt{B+T} section 7.1).
We use an instantaneous hyperbolic orbit orbit approximation rather than the impulse approximation \citep{spitzer58} (the high velocity
limit of the hyperbolic orbit approximation) because the E2 encounter velocity is oriented nearly perpendicular
to the disc.   When the perturber velocity vector is perpendicular to the disc, 
the impulse approximation gives no  vertical velocity perturbation to stars in the disc, making it impossible to 
explore the role of  vertical epicyle phase wrapping.  However,  the more accurate hyperbolic orbit
approximation gives a small velocity perturbation in the direction opposite to the velocity vector and so there
is a vertical velocity perturbation.
 
A low mass particle encountering mass $M$ with a relative velocity ${\bf V}$ and with impact parameter $b$
on a hyperbolic orbit has, after the encounter, a change in velocity
\begin{eqnarray}
 \Delta v_\perp &=& \frac{2bV^3}{GM} \left(1 + \frac{b^2 V^4}{G^2 M^2} \right)^{-1} \label{eqn:vperp} \\
 \Delta v_\parallel &=& {2V} \left(1 + \frac{b^2 V^4}{G^2 M^2} \right)^{-1} \label{eqn:vpar}
 \end{eqnarray}
 (\citealt{B+T} section 7.1)
 where $\Delta v_\perp$ is in the direction toward position of closest approach and $\Delta v_\parallel$
 is in the opposite direction of the relative velocity vector of $M$ and the low mass particle.
 
To compute the impact parameter and vertical and parallel direction vectors
 we consider linear trajectories for both star and dwarf galaxy,
${\bf x}_d(t) = {\bf x}_{d0} + {\bf v}_d (t - t_0) $
for the dwarf galaxy with position and velocity ${\bf x}_{d0}$ and ${\bf v}_d$ at time $t_0$ when
it passes through the Galactic plane.
For a star at ${\bf x}_{s0}$ and velocity ${\bf v}_s$ a linear trajectory
${\bf x}_s(t) = {\bf x}_{s0} + {\bf v}_s (t - t_0). $
These linear trajectories are those in the vicinity of the encounter, 
ignoring the gravitational interaction during encounter
and the background Galactic potential, and we only use them to compute the velocity perturbations.
In the frame with relative velocity 
${\bf V} = {\bf v}_d - {\bf v}_s$
centered on the dwarf galaxy,
 the linear trajectory of the star is
$ {\bf x}_{s,com} = {\bf x}_{s0} - {\bf x}_{d0} - {\bf V} (t - t_0). $
By minimizing distance to the dwarf  we find that 
the star (on the linear trajectory) is closest to the dwarf galaxy at time 
$t_{min} = t_0 + {\bf V} \cdot ({\bf x}_{s0} - {\bf x}_{d0}) V^{-2} .$
At this time the star  is at  
$${\bf x}_{s,com} (t_{min}) = {\bf x}_{s0}  - {\bf x}_{d0} - {\bf V} \left[ {\bf V} \cdot ({\bf x}_{s0} - {\bf x}_{d0}) \right] V^{-2} .$$
The vector between dwarf and star (and pointing toward the dwarf for the linear trajectory at closest approach)
is the equivalent to this but with opposite direction
$${\bf b} = {\bf x}_{d0}  - {\bf x}_{s0} + {\bf V} \left[ {\bf V} \cdot ({\bf x}_{s0} - {\bf x}_{d0}) \right] V^{-2} $$
The length of ${\bf b}$ is the impact parameter, $b$.  The direction of ${\bf b}$ 
gives the perpendicular direction for the velocity change (equation \ref{eqn:vperp}).
The parallel velocity perturbation (equation \ref{eqn:vpar}) is in the direction of the relative
velocity vector $- \hat {\bf V}$ with unit vector
$\hat {\bf V} = ({\bf v}_d - {\bf v}_s)/|({\bf v}_d - {\bf v}_s)|.$
Rewriting equations \ref{eqn:vperp} and \ref{eqn:vpar},
we can write the total velocity perturbation of the star as 
\begin{equation}
\begin{split}
 \Delta {\bf v} =  \frac{2bV^3}{GM(b)} \hat {\bf b} \left(1 + \frac{b^2 V^4}{G^2 M(b)^2} \right)^{-1} \\
-{2V} \hat {\bf V} \left(1 + \frac{b^2 V^4}{G^2 M(b)^2} \right)^{-1} 
\end{split}
\label{eqn:tot_v}
\end{equation}
with first term the perpendicular component and the second term the parallel component.
Here $\hat {\bf b}$ is the unit vector with the direction of ${\bf b}$.
We approximate the dwarf galaxy mass as that integrated out to the impact parameter, $M(b)$ to
take into account the spatial extent of the  dwarf galaxy and to limit the size of the perturbations
at small impact parameters.  A Hernquist mass model (mass interior to radius $r$)
\begin{equation}
M(r) = M_d \frac{r^2}{(r+a_H)^2} \label{eqn:Mhern}
\end{equation}
is used to model mass distribution of the dwarf galaxy, with Hernquist scale length $a_H = 3$ kpc.
The scale length of the dwarf galaxy truncates the size of the largest velocity perturbations.  However, as
large perturbations
are only induced over a very small area, the velocities for most of the perturbed disc are insensitive to the 
value of $a_H$ used.

We restrict the effect of the hyperbolic approximation to stars within 18 kpc of the location of impact (the
positions given in Table \ref{tab:dwarf}). 
Impact parameters larger than this (such as those for stars on the opposite side of the disc) would not be
strongly perturbed by the encounter and the perturbations would not be well approximated by the 
instantaneous hyperbolic orbit approximation.

In the limit of high velocity or $ b^2 V^4/(GM(b))^2 \gg 1$ equation \ref{eqn:tot_v} becomes
\begin{equation}
\Delta {\bf  v} = \hat {\bf v} \frac{2 GM(b)}{bV} \label{eqn:impulse}
 \end{equation}
 resembling the impulse approximation for a point mass of mass $M(b)$.
For large impact parameter, $b\gg a_H$,  the enclosed mass $M(b) \sim M_d$ and 
$$ \Delta {\bf v} \approx \hat {\bf b}  \frac{2GM_d}{bV} $$ as would be expected for the impulse
approximation for a point mass of mass $M_d$. 
The  impulse approximation \citep{spitzer58,cincotta91} adopts a straight line for the
 the trajectory of the perturber (here the dwarf galaxy), with respect to a star in the disk.
 The velocity perturbation of the star caused by an extended potential perturber (rather than a point mass) can be computed as
\begin{equation}
\Delta {\bf v} \approx \int_{-\infty}^\infty dt \nabla \Phi(t)  \label{eqn:pot}
\end{equation}
\citep{cincotta91}
where $\Phi (t)$ is the potential of the perturber on a linear trajectory.  This can be integrated analytically
for the Plummer model \citep{cincotta91}.  Using the Hernquist gravitational potential, 
$\Phi(r) = -GM_d/(r+a_H)$, consistent with equation \ref{eqn:Mhern},
  equation \ref{eqn:pot} gives 
\begin{equation}
 \Delta{\bf v} \approx {\hat {\bf b}} \frac{2GM_d }{V b} g(a_H/b) 
 \label{eqn:impulsehern} \end{equation}
with function
$$g(u) =   \int_{0}^\infty \frac{dx}{ (\sqrt{x^2 + 1} + u)^2  \sqrt{x^2 + 1}}. $$ 
In the limit of $b \gg a_H$,  the limit $\lim_{u\to 0} g(u) = 1$ and we recover the impulse approximation for point mass $M = M_d$, consistent 
with the total enclosed perturber mass at large radius.    

How accurate is equation \ref{eqn:impulse} compared to equation \ref{eqn:impulsehern}?
Computing the ratio of the two expressions at $b=a_H$ we find that equation \ref{eqn:impulse}
underestimates $|\Delta {\bf v}|$ by 30\%.  This underestimate is due to the neglect of the mass outside of radius $r=b$
in equation \ref{eqn:impulse}.   Drawing from this computation, we estimate that we underestimate
the velocity perturbation at $b=a_H$ using the hyperbolic orbit approximation (equation \ref{eqn:tot_v}) by about the same fraction.
For our two encounters most stars have impact parameters larger than $a_H$ so we do not expect
this underestimate to significantly influence our results.

\end{document}